\documentclass[review]{elsarticle}

\usepackage[utf8]{inputenc}
\usepackage{amsmath}
\usepackage{amssymb}
\usepackage{graphicx}
\usepackage{dirtytalk}
\usepackage{charter}    
\usepackage{subcaption}   
\usepackage{hyperref}
\usepackage[acronym]{glossaries}
\usepackage{bm}
\usepackage{fullpage} 
\hypersetup{
    colorlinks=true,
    linkcolor=blue,
    filecolor=magenta,      
    urlcolor=blue,
    bookmarks=true,
    citecolor=blue,
} 
\graphicspath{ {./} }

\def\Runo{{\cal R}_{\rm E}}
\def\Rdue{{\cal R}_{\rm D}}
\def\Rtre{{\cal R}_{\rm c}}
\def\nl{n_{\ell}}
\def\I{{\bf I}}
\def\<{\langle}
\def\>{\rangle}
\def\x{{\bf x}}
\def\D{{\cal D}}
\def\M{{\cal M}}
\def\T{\bm{\theta}}
\def\traza{{\rm tr\,}}
\newcommand{\1}{1}

\newacronym{s}{{\rm s}}{shape}
\newacronym{a}{{\rm t}}{texture}

\begin{document}

\begin{frontmatter}

\title{Information-theoretical analysis of the neural code for decoupled face representation}


\author[firstaddress]{Miguel Ib{\'a}{\~nez}-Berganza}
\author[secondaddress]{Carlo Lucibello}
\author[thirdaddress]{Luca Mariani}
\author[fourthaddress]{Giovanni Pezzulo\corref{mycorrespondingauthor}}
\cortext[mycorrespondingauthor]{Corresponding author}
\ead{giovanni.pezzulo@istc.cnr.it}

\address[firstaddress]{Istituto Italiano di Tecnologia. Largo Barsanti e Matteucci, 53, 80125 Napoli, Italy; Sapienza University of Rome. Piazzale Aldo Moro, 5. 00185 Roma, Italy}
\address[secondaddress]{Institute for Data Science and Analytics, Bocconi University. Via Roberto Sarfatti, 25, 20100 Milano, Italy}
\address[thirdaddress]{Department of Physics ``E. R. Caianiello'', University of Salerno. Via Giovanni Paolo II 132, 84014, Fisciano, Italy}
\address[fourthaddress]{Institute of Cognitive Sciences and Technologies, National Research Council, Via San Martino della Battaglia 44, 00185 Rome, Italy}


\begin{abstract}

	Processing faces accurately and efficiently is a key capability of humans and other animals that engage in sophisticated social tasks. Recent studies reported a decoupled coding for faces in the primate inferotemporal cortex, with two separate neural populations coding for the geometric position of (texture-free) facial landmarks and for the image texture at fixed landmark positions, respectively. Here, we formally assess the efficiency of this decoupled coding by appealing to the information-theoretic notion of description length, which quantifies the amount of information that is saved when encoding novel facial images, with a given precision. We show that despite decoupled coding describes the facial images in terms of two sets of principal components (of landmark shape and image texture), it is more efficient (i.e., yields more information compression) than the encoding in terms of the image principal components only, which corresponds to the widely used eigenface method. The advantage of decoupled coding over eigenface coding increases with image resolution and is especially prominent when coding variants of training set images that only differ in facial expressions. Moreover, we demonstrate that decoupled coding entails better performance in three different tasks: the representation of facial images, the (daydream) sampling of novel facial images, and the recognition of facial identities and gender. In summary, our study provides a first principle perspective on the efficiency and accuracy of the decoupled coding of facial stimuli reported in the primate inferotemporal cortex.
	
\end{abstract}

\end{frontmatter}


\section{Introduction \label{intro}}

Recognizing faces and facial expressions with high accuracy is central for many cognitive and social tasks that primates (and possibly other animals) perform every day \cite{leopold2010comparative}. Several studies reported single neurons in the ventral visual stream -- and particularly in the so-called ``face patches'' of the inferotemporal (IT) cortex -- that are exquisitely sensitives to faces \cite{tsao2006cortical,tsao2008mechanisms}. 

A recent landmark study greatly contributed to shed light on the neural code for facial identity in the IT of macaques \cite{chang2017}. This study reported that faces might be represented as feature vectors in a relatively low-dimensional ($\sim$50D) {\it face space} \cite{valentine2016}, with IT neurons tuned to single axes of variation of the face space and insensitive to changes in other, orthogonal axes\footnote{
Indeed, neurons are believed to encode principal components {\it linearly} but not necessarily one-to-one, see \cite{chang2017}. In particular, if $\bf y$ is the vector of neurons' normalized firing rates and ${\bf x}'$ is the vector of principal components in the face space, an orthogonal matrix $O$ relates $\bf y$ and ${\bf x}'$: ${\bf y} = O {\bf x'}$.}. Interestingly, distinct subpopulations of neurons appear to project faces onto two distinct sets of axes, which encode the geometric {\it shape} of a face and its {\it texture} (at a fixed shape) separately. The shape coordinates describe the main facial proportions, whereas the texture coordinates bring information about the detailed form of facial soft tissues, the skin texture and tonality, and cues to the facial shape in the depth dimension, through the light reflection.

From a computational perspective, these findings suggest that the IT cortex might form a generative model in which shape- and texture-related information is \emph{decoupled} into separate factors (aka disentangled or factorised). The resulting \emph{decoupled coding} ($\Rdue$) resembles closely a computer vision model called the Active Appearance Model (AAM) \cite{edwards1998interpreting}. A recent computational study indicates that the AAM provides a very good fit for the single cell IT data of \cite{chang2017}, outperforming most standard deep network models of visual processing in the ventral stream \cite{chang2021explaining}. While the deep networks achieve a high score in face (or object) recognition, they do so by \emph{multiplexing} the same information into different neurons, which is the opposite of the decoupling strategy reported in IT neurons by \cite{chang2017}. In keeping, another computational study \cite{higgins2021} showed that using a deep generative model ($\beta$-VAE), with the explicit objective of disentangling facial images into separate latent factors, provides a good account of IT neural firings \cite{chang2017}. 


In a series of neural network simulations \cite{chang2017,chang2021explaining}, the face processing performance of the decoupled coding ($\Rdue$) that emerges from IT recordings is compared with a simpler scheme that is standard in computer vision: \emph{eigenface} ($\Runo$) \emph{coding}  \cite{sirovichLowdimensionalProcedureCharacterization1987,Turk:1991:ER:1326887.1326894,valentine2016}. In both $\Runo$ and $\Rdue$ codings, each neuron ``projects'' facial images linearly onto one axis of variation of the face space. However, the projection is different in the two. In $\Runo$ coding, the neurons simply encode the projection of the input face into the axes of variation of the original set of known facial images. Rather, in $\Rdue$ coding, the input facial image is first divided into two sources of information: a (shape-free) average-shaped or {\it uniformed} facial image whose texture corresponds to that of the input face, and a (texture-free) vector of Cartesian coordinates of some facial reference points called {\it landmarks}, describing the input face shape. Then, in $\Rdue$ coding, one set of neurons encodes the linear projection of the input {\it uniformed} facial image on the axes of variation {\it of uniformed images}, whereas another set of neurons encodes the projection of the input vector of landmark coordinates on the axes of variation of vectors of landmark coordinates. As reported in \cite{chang2017}, the decoupled coding scheme $\Rdue$ explains a significantly higher fraction of neural data variance than the $\Runo$ coding. 

While the above studies assess that decoupling information is a key ingredient of facial processing in primates, it is still unclear why this is the case. A plausible formal rationale for the decoupling of shape and texture parameters (as done in the AAM and related models) is that they might vary independently in real life conditions. For example, small variations in facial expressions entail a significant change of shape but not texture, whereas different conditions of luminosity and age may induce significant variations in texture but not shape \cite{chang2017}. This line of reasoning leads to the untested idea that decoupled coding entails not just a more \emph{accurate} but also a more \emph{efficient} (or compact) description of facial data. 

Indeed, various normative principles have been proposed that characterize the \emph{efficient coding} of data from a source in terms of information demands \cite{attneave1954some,barlow1961possible,schneidman2003} (see references in \cite{fairhall2012}). From an information theoretic perspective, a formal measure of code efficiency is its {\it description length}: the best model is the one that minimizes the amount of information (bits) required to encode both the data, in terms of the model's latent variables, and the model parameters themselves \cite{rissanen1978modeling,rissanen1999hypothesis,mackay2003}. This implies that a more complex model, which has more free parameters and requires more memory to be encoded, will only outperform a simpler model if it affords significantly more data compression -- which in turn requires that it captures well the statistical structure of the data.

Here we use the notion of efficient coding to ask whether, why, and in which conditions the neural code for face representation found in monkey IT neurons, which is based on texture-shape decoupling ($\Rdue$ coding), is more efficient than a simpler description in terms of principal components of facial images, without texture-shape decoupling (eigenface coding $\Runo$). For this, we compare the description length of (the principal components of) the two elements of $\Rdue$ coding -- namely, shape-free texture and shape coordinates -- with the description length of (the principal components of) the original facial images, using the same stimuli dataset as in the monkey study of \cite{chang2017}. To preview our results, we show that the neural code based on texture-shape decoupling ($\Rdue$) is more efficient than the eigenface coding ($\Runo$). This is because storing the principal components of few (significant) landmark coordinates comes at the cost of {\it little extra information}, but it confers the advantage of uniforming (shape-free) facial images. In turn, the uniformed facial images are significantly more correlated than the original set of facial images and can be described using fewer principal components -- hence yielding an overall positive information gain. In keeping, our results reveal that the advantage of decoupled coding increases with image resolution and when encoding variants of training set images that differ for facial expressions. This result is interesting, as it shows that the decoupled coding is most effective in a conditions that is frequent in social cognitive tasks, such as the identification of changes of expression or age in known faces. Finally, to further consolidate our findings, we show that decoupled coding leads to a higher efficiency in a range of cognitively relevant tasks, which include the daydream generation of novel faces, the synthesis of unknown faces, and the recognition of facial identities and gender.


\section{Materials and methods \label{materials}}

\subsection{Database}

In our analysis, we use the FEI database \cite{thomazNewRankingMethod2010,fei}, which was also used in the characterisation of the neural code of facial identity in macaques \cite{chang2017}. The FEI database comprises $N=400$ b/w pictures of dimension $w_{\rm max}\times h_{\rm max}=250\times 300$ pixels, accompanied by the spatial coordinates of $\nl=46$ standard landmarks for each image. 

\subsection{Texture and shape coordinates}

Let the training set consist of $N_{\rm tr}$ facial images, ${\cal I}=\{\I(n)\}_{n=1}^{N_{\rm tr}}$, where $\I(n)$ is the $n$-th image, and of $N_{\rm tr}$ vectors of shape coordinates ${\cal L}=\{{\bm\ell}(n)\}_{n=1}^{N_{\rm tr}}$, where ${\bm\ell}(n)$ is the vector of shape coordinates characterising the geometry of the $n$-th facial image. All images are vectors $\I(n)=(I_1(n),\ldots,I_{d_{\rm t}}(n))$ of dimension $d_{\rm t}=w\times h$, where $w,h$ are the width and height of the images in pixels (grid spacing units). The ${\bm \ell}$-vector components are the $\sf x$ or $\sf y$ Cartesian coordinates of $\nl$ representative landmarks of the $n$-th facial image: ${\bm \ell}(n)=(\ell_1(n),\ldots,\ell_{d_{\rm s}}(n))$, with $d_{\rm s}=2\nl$.

\subsection{Formal definitions of eigenface coding ($\Runo$) and decoupled  coding ($\Rdue$)}

\begin{figure}
\begin{center}
    \makebox[\textwidth][c]{
    \includegraphics[width=0.14\textwidth]{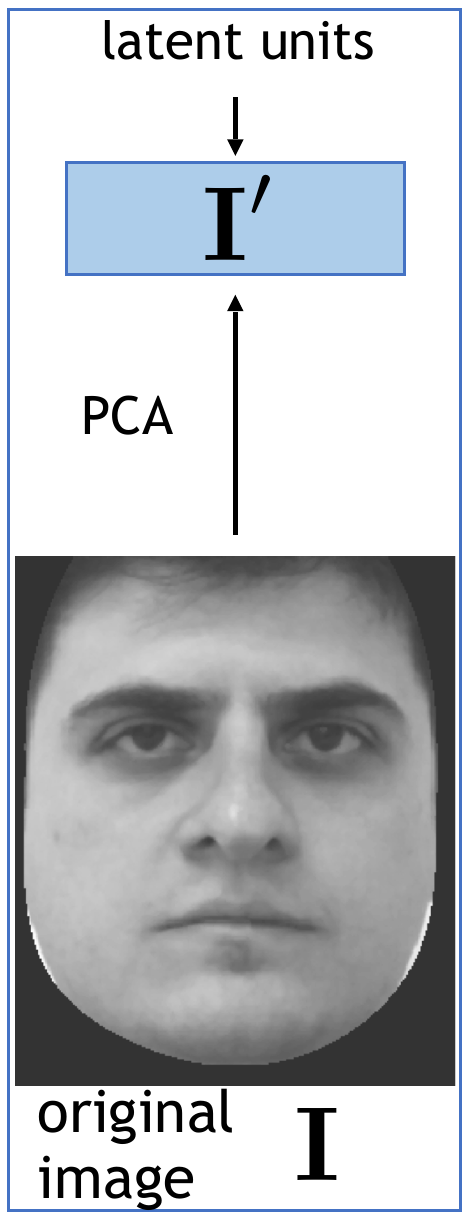} \hspace{1cm}  \includegraphics[width=0.45\textwidth]{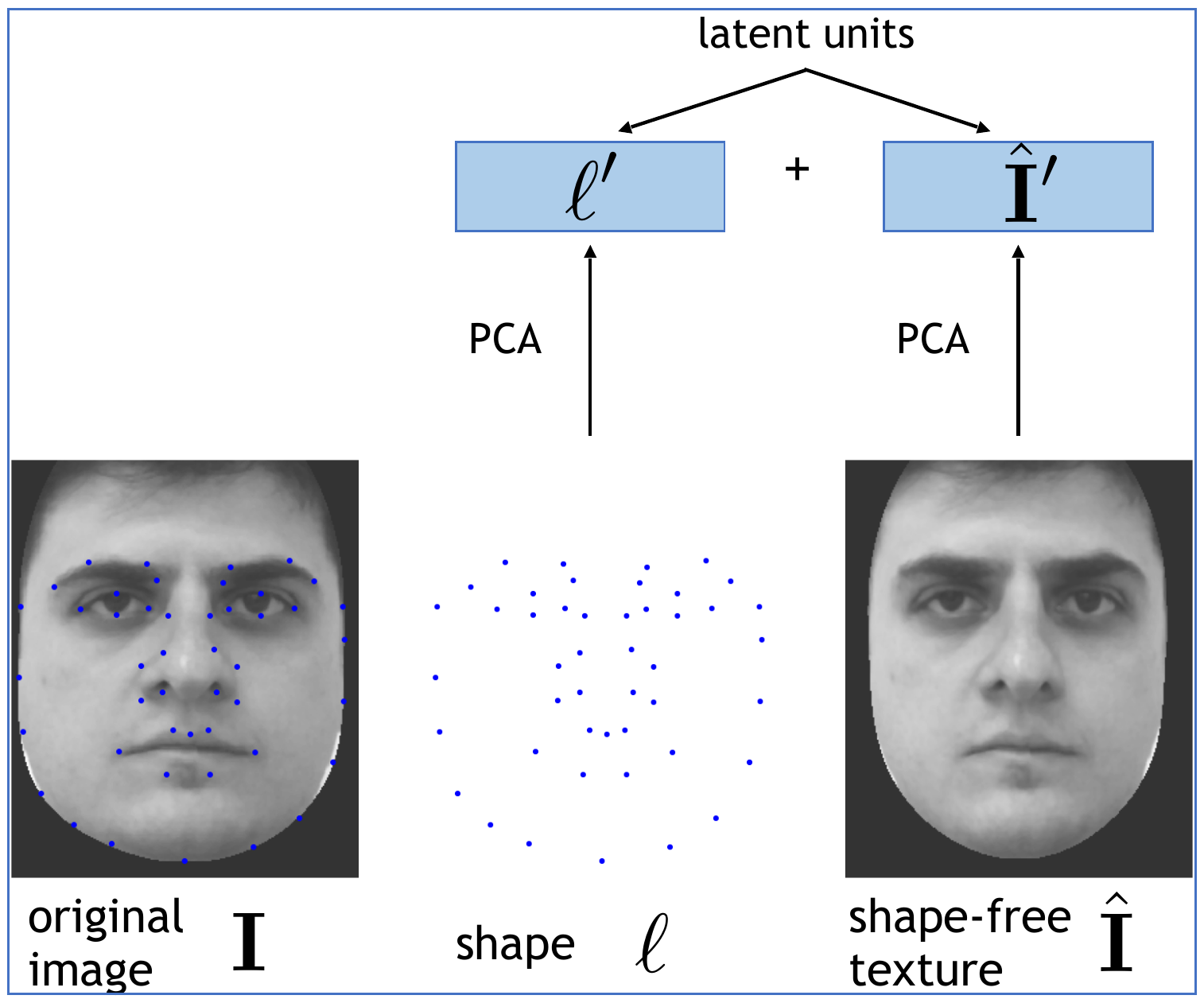}}
	\caption{Schematic illustration of eigenface coding ($\Runo$, left) and decoupled coding ($\Rdue$, right). See the main text for explanation.}
    \label{fig:coordinates}
\end{center}   
\end{figure}

We consider two alternative neural codes for facial images: \emph{eigenface} ($\Runo$) \emph{coding} and \emph{decoupled} ($\Rdue$) \emph{coding}; see Table \ref{table1} and Figure \ref{fig:coordinates}. Both the $\Rdue$ and the $\Runo$ codings represent facial images in terms of Principal Components (PCs), but over different facial coordinates of the training set (i.e., over different datasets). Specifically, they represent a generic image $\I$ as follows:

\begin{enumerate}
	\item \emph{Eigenface coding} ($\Runo$) represents the image in terms of its PCs, $\I'$. In mathematical terms, $\I' = E_p^{\rm (E)} \cdot \I$, where $E_p^{\rm (E)}$ is the $p\times d_{\rm t}$ matrix composed of the first $p$ (row) eigenvectors of the unbiased estimator of the correlation matrix $C$ of training-set images, $C_{ij} = \<I_i(n) I_j(n)\>$, where $\<\cdot\>=(1/N_{\rm tr}\sum_n \cdot)$ is the empirical average over the training-set, and where all the vector components are null-averaged, $\<x_i\>=0$. This representation does not make use of the shape coordinates.
	\item \emph{Decoupled coding} ($\Rdue$) represents the image in terms of two sets of PCs, one for shape and one for texture facial coordinates. To obtain these coordinates, each original image $\I(n)$ in the  training set is first deformed by means of image-deformation algorithms (see \cite{image_deformation, ibanez2019,ibanez2020} and the Supporting information for details), in such a way that its landmark coordinates ${\bm \ell}(n)$ are dragged to the {\it average position of the landmark coordinates in the training-dataset}, and that the rest of the image pixels are deformed coherently (so that the resulting facial image is as much realistic as possible). The resulting image will be called the {\it uniformed} image $\hat\I(n)$ (see figure \ref{fig:coordinates}). We refer to {\it uniformed texture  coordinates}, or simply {\it texture  coordinates}, as the {\it uniformed} (shape-free) image coordinates $\hat \I$, of an image $\I$ given $\bm \ell$ (and the average position of the landmarks $\<{\bm \ell}\>={\bf 0}$). This procedure permits decoupling the original dataset in two datasets of coordinates: the (texture-free) shape coordinates $\cal L$ and the  (shape-free) uniformed images $\hat{\cal I}=\{\hat \I(n)\}_{n=1}^{N_{\rm tr}}$. 

The novel image $\I$ to be represented is then decomposed in PCs in texture and shape spaces separately, $\hat{\I}'=E_p^{({\rm t})}\cdot\hat \I$, ${\bm\ell}'=E_p^{({\rm s})}\cdot{\bm \ell}$, where $E_p^{({\rm t})}$ are the eigenvectors of $C^{({\rm t})}_{ij}=\<\hat I_i(n) \hat I_j(n) \>$, and $E_p^{({\rm s})}$ those of $C^{({\rm s})}_{ij}=\<\ell_i(n)\ell_j(n)\>$.
\end{enumerate}

\begin{table}[!ht]
\caption{\bf Outline of the two alternative schemes for the representation of facial images: eigenface coding ($\Runo$) and decoupled coding ($\Rdue$).\label{table1}}
\begin{tabular}{l|ccccccc} 
	Code & \\
	\hline
	$\Runo$ & $(\I,{\bm\ell})$ &  &  & $\begin{subarray}{c} {\rm projection} \\ \to \end{subarray}$ & $\I'_p=E^{\rm (E)}_p\cdot \I$ & $\to$  & $(\I'_{p_{\rm s}})$ \\
	\hline
	$\Rdue$ & $(\I,{\bm\ell})$ & $\begin{subarray}{c} {\rm uniformation} \\ \to \end{subarray}$ &  $(\hat\I,{\bm \ell})$ & $\begin{subarray}{c} {\rm projection} \\ \to \end{subarray}$ & $\begin{subarray}{c} \hat\I'_p=E^{({\rm t})}_p\cdot \hat\I \\ {\bm\ell}'_p=E^{({\rm s})}_p\cdot {\bm \ell}\end{subarray}$  & $\to$ & $(\hat\I'_{p_{\rm s}},{\bm\ell}'_{p_{\rm t}})$\\
	\hline
\end{tabular}
\end{table}

\subsection{Description length analysis}
\label{sec:description_length}

\subsubsection{Intuition behind the description length analysis}


The Principal Component Analysis representation of a given set of coordinates in the face space with $p$ principal components ($p$-PCA) can be viewed both as a generative model, inducing a Gaussian distribution over facial coordinates, and as a form of data compression and dimensionality reduction \cite{beyeler2019}. These two aspects are naturally linked by the notion of {\it description length} \cite{mackay2003}. PCA is a form of dimensionality reduction, since it describes each $d$-dimensional vector $\x$ as a shorter, $p$-dimensional vector $\x'_p=E_p\cdot \x$. In turn, this implies a {\it compression} ability. Consider a dataset in which each coordinate $x_i$ (say, each pixel value, if the vectors $\x$ are images) varies uniformly in a range $R$. In the absence of any prior knowledge regarding the dataset content, the amount of information per sample and coordinate needed to store the raw dataset with precision $\epsilon$ per coordinate is simply $l_0=\log_2 (R/\epsilon)$ bits. Normally, the information needed to store the $p$ principal components of each vector of the dataset $\D'=\{\x'_p(s)\}_{s=1}^N$ is lower than $l_0$, even if $p=d$. Indeed, if the dataset exhibits significant pairwise correlations between couples of variables, many principal components will exhibit a variance $\lambda_i$ lower than the average variance $R^2/12$, and they will consequently require fewer bits to be stored. 

The amount of information necessary to encode a dataset $\cal D$ in terms of (the latent variables of) a probabilistic model $\M$ of the dataset vectors is called {\it description length}, $L_\M({\D})$. Crucially, the description length is formally related to the Bayesian data evidence, or joint marginal likelihood of the dataset $\cal D$ according to the model $\M$ in the following way: $L_\M(\D) = -\log_2 [P_{\cal M}({\D}) |\epsilon|^d]$, where $P_{\cal M}(\D)$ is the data evidence according to $\M$ and $\epsilon$ is the precision per coordinate with which the dataset should be described. Description length is therefore equivalent to -- and provides an information-theoretic interpretation of -- Bayesian model evidence. The value of $p$ for which the dataset presents a higher Bayesian evidence is the one presenting an optimal accuracy/complexity trade-off and, consequently, the one presenting a lower description length. In other words, description length analysis evaluates the efficiency of a particular code, taking into account both its accuracy and its complexity. In this perspective, a good code is the one that does not employ too much information to describe a given input with a given tolerance. Indeed, the model that presents lower description length at fixed precision, is also the one that manages to describe the dataset with a smaller error, $\log_2 \epsilon = -\log_2 P_{\cal M}(\D) -L$, when the amount $L$ of available storing information is fixed.

In the case that we study here, the model $\M$ is $p$-PCA and the explicit expression for $P(\D)$ is interpretable \cite{mackay2003}. The description length may be decomposed in two terms, $L(\D) = {\sf S(\D|\T^*)} + {\sf O}(\T^*)$, that we will call the {\it empirical entropy} ${\sf S(\D|\T^*)}$ and the {\it Occam length} ${\sf O}(\T^*)$.\footnote{$\T^*$ are the model parameters (the eigenvector matrix $E_p$ and the vector of averages $\bm \mu$) fitted as the Maximum Likelihood value for a training set $\D_{\rm tr}$, that may be different from $\D$.} These two terms may be interpreted as the amount of information needed to encode (without losses\footnote{Crucially, the $p$-PCA model induces a (Gaussian) probabilistic model defined on the $d$-dimensional linear space of the data also when $p<d$. The $d-p$-dimensional subspace not expanded by the $p$ fitted empirical eigenvectors (corresponding to the $p$ largest eigenvalues $\lambda_j$) is described with a constant, degenerated noise eigenvalue $\bar\lambda$. As a consequence, despite the PCA representation of vectors in terms of $p$ principal components {\it is a lossy representation} for $p<d$, the induced likelihood and empirical entropy $\sf S$ terms in the description length of each vector do {\it take into account the losses}.}) the dataset $\D$ in terms of $p$ principal components, and the model parameters $\T^*$ once for all the vectors (that are needed to recover each vector $\x$ from its principal components $\x'$), respectively. When increasing the number of model parameters $p$, the empirical entropy of the training dataset decreases, but the Occam length generally increases, since more eigenvectors $E_p$ must be stored -- and they must be stored with a higher precision. Overfitting occurs when this balance is no longer worth, and the description length increases for increasing $p$. 

\subsubsection{Definition of the \emph{information gap} criterion for the comparison of decoupled coding and eigenface coding}

We measured the description length (in bits) of the alternative coding schemes $\Runo$ and $\Rdue$ of facial images $\I$ that belong to a set of known images that have been used to train the model (training set) and to a set of unknown images that have not been used to train the model (test set). 

Eigenface coding ($\Runo$) encodes the original images $\I$ in terms of their principal components. We denote the description length associated to the compression of a dataset ${\cal I}$ according to $\Runo$ as $L_{{\cal I}_{\rm tr},p}({\cal I})$. The two sub-indices of $L$ specify the model; they are, respectively, the training set with which the model parameters have been trained\footnote{PCA induces a multivariate normal distribution whose average vector $\bm \mu$ is the unbiased estimator of this quantity in the dataset $\cal I_{\rm tr}$, and whose covariance matrix $C$ shares the $p$ largest-eigenvalues and corresponding eigenvectors with the sample covariance matrix of the set $\cal I_{\rm tr}$ (see the details in the Supplementary information).}, and the value of $p$. This information completely determines the $p$-PCA model. 

Decoupled coding $\Rdue$ describes each facial image $\I$ in terms of the principal components of (shape-free) uniformed texture and (texture-free) landmark coordinates $\hat\I'$, ${\bm\ell}'$. This decoupling is motivated by the hypothesis that shape-free uniformed images can be compressed more easily compared to the original images and their unknown variants; or in other words, that there is a parsimonious description of the dataset of uniformed faces $\hat{\cal I}$, when represented in terms of their principal components $\hat{\cal I}'$. However, since decoupled coding ($\Rdue$) exploits both the the texture and the shape coordinates ($\bm \ell$ and $\hat \I$) of each facial vector, it has to store both sets of principal components ($\bm \ell'$ and $\hat \I'$) to represent the original image. Moreover, it has to store the principal axes in both the space of texture and shape coordinates\footnote{Eigenface coding is usually understood as a representation in terms of non-local principal components. Please note that it would be misleading to interpret the decoupled coding $\Rdue$ as being local instead, just because it uses landmark coordinates. Indeed, $\Runo$ represents facial images in terms of principal components ${\I}'$, and so does $\Rdue$: $\hat{\I}'$ are non-local in the sense that each component is a linear combination of pixel intensities occupying different positions in the image canvas; ${\bm\ell}'$ is non-local as well, in the sense that each component is a linear combination of different landmarks' Cartesian coordinates.}. 

The key question addressed here is whether the extra information cost required to store shape coordinates might be compensated by the smaller cost to store the {\it uniformed} texture principal components $\hat \I'$. In principle, the uniformed set of images $\hat \I$ might be compressed more easily, given that the inhomogeneities induced by the difference in landmark positions have been removed from the dataset -- at least, if the resolution of the image is large enough. This implies that encoding the uniformed images could in principle require a smaller number of PCs without loss of precision, with respect to the set of raw images.


To quantify the difference in description length between uniformed texture coordinates, non-uniformed texture coordinates, and shape coordinates, we define a summary measure that we call an {\it information gap} and which jointly considers two factors. The former factor ($\mathcal G_{1}$), the {\it texture information gap}, accounts for the difference in the description lengths of the non-uniformed and uniformed image datasets:

\begin{eqnarray}
\label{eq:gain}
 {\mathcal G_{1}} = L_{{\cal I}_{\rm tr},p}({\cal I}) &-& L_{\hat{\cal I}_{\rm tr},\hat{p}}(\hat{\cal I}) \\
 \begin{subarray}{l} \textrm{bits to compress the database} \\ \textrm{of non-uniformed images } {\cal I} \end{subarray} &-& 
	 \begin{subarray}{l} \textrm{bits to compress the database} \\ \textrm{of uniformed images } \hat{\cal I} \end{subarray}   \nonumber
\end{eqnarray}
where $\hat {\cal I}$ is the dataset composed by the uniformed facial images in $\cal I$. Note that in both the description length terms, the model $\M$ is assumed to be $p$-PCA. In these equations, $p$ may be taken as the optimal value according to Bayesian model selection, i.e., the value (say, $p^*$) for which the description length of $\cal I$ is minimum, and the same for $\hat p^*$. 

The latter factor ($\mathcal G_{2}$) is the description length of the set of shape coordinates ${\cal L}=\{{\bm \ell}(n)\}_n$:

\begin{eqnarray}
\label{eq:dlength}
{\mathcal G_{2}} = L_{{\cal L}_{\rm tr},p}(\cal L)
\end{eqnarray}
where $\cal L$ is the set of landmarks corresponding to the images $\cal I$, and ${\cal L}_{\rm tr}$ to those in ${\cal I}_{\rm tr}$. 

The information gap combines these two factors ($\mathcal G = \mathcal G_{1} - \mathcal G_{2} $) and measures the efficiency (in information-theoretic terms) of decoupled coding $\Rdue$ compared to eigenface coding $\Runo$:

\begin{eqnarray}
\label{eq:conditionongain}
 {\rm Information\ gap\ in\ favour\ of\ \Rdue:\ } {\mathcal G} = {\mathcal G}_1 - {\mathcal G}_2 
\end{eqnarray}

Decoupled coding can be considered more efficient if the information gap $\mathcal G$ is greater than zero. In other words, given a dataset of facial images, the representation of $\Rdue$ is more efficient than that of $\Runo$ to the extent that it provides a more accurate description of the dataset, using the same amount of available information (see also the Supporting information).

Note that in practice, our comparison boils down to computing and comparing the description length of {\it different datasets}: $\hat {\cal I}$, ${\cal I}$, $\cal L$ (the first one being computed from the last two), in terms of their principal components, i.e., using the same statistical model ($p$-PCA) for all the three datasets, {\it but with different covariance matrices} (inferred from $\hat {\cal I}_{\rm tr}$, ${\cal I}_{\rm tr}$, ${\cal L}_{\rm tr}$) {\it and values of $p$}. We can therefore assess the efficiency of decoupled coding $\Rdue$ by considering the difference in description length between ${\cal I}$ and $\hat {\cal I}$ -- and whether it compensates for the description length of $\cal L$\footnote{Note that this approach is different from a standard {\it Bayesian model selection}, in which different models are compared using the same dataset; see more details in the Supplementary information, section {\it Relation with Bayesian Model Selection.}}.


\subsubsection{Precision of shape and texture coordinates and estimation of the description length}

For the calculation of the description length $L_{{\cal D}_{\rm tr},p}({\cal D})$ we exploit the analytical solution of the Bayesian evidence of a multivariate normal distribution \cite{minka2000}. Note that in the case of texture coordinates, that are strongly undersampled $N\ll d_{{\rm t}}$, it is essential to use such (asymptotically) exact expression, instead of its more common Bayesian Information Criterion approximation \cite{mackay2003,Myung2000}, see the Supporting information. The training dataset and the number of principal components completely define the parameters of the inferred normal distribution ${\cal M}=({\cal D}_{\rm tr},p)$, whose Bayesian evidence $P_{{\cal M}}({\cal D})$ can be estimated analytically (see \cite{minka2000} and the formulae in the Supplementary information). The estimation of the description length depends, as said before, on a precision per coordinate $\epsilon$ with which the data should be described by the probabilistic model, $L_{\cal M}({\cal D}) = -\log_2(P_{{\cal M}}({\cal D})) - d \log_2(\epsilon)$. The second term in this equation is equivalent to the factor transforming the differential entropy of a continuous probability distribution into a genuine entropy \cite{jaynes1957,jaynes1983brandeis,caticha2021} once the continuous variables have been discretised in bins of of size $\epsilon$. 

In the case of texture coordinates, in which the vectors are images and the vector components $I_i$ are 8-bit grayscale values in the range $(0,255)$, the natural cutoff value is $\epsilon_{\rm t}=1$. In the case of shape coordinates, the coordinates $\ell_i$ are grid integers varying in a range $[1,h]$, so that the natural choice should be $\epsilon_{\rm s} = 1$ for the largest resolution $h_{\rm max}$, and $h_{\rm max}/h$ for lower resolutions $h<h_{\rm max}$ (since we change the coordinates resolution by scaling the largest-resolution coordinates as $\ell_i\to(h/h_{\rm max})\ell_i$). Scaling the precision in this way, the empirical entropies of shape coordinates do not depend on the resolution (see also {\it Likelihood and evidence of shape coordinates} in the Supplementary information). We actually intentionally underestimate the shape coordinates' cutoff and set $\epsilon_{\rm s} = 0.1 (h_\text{max}/h)$, so that we overestimate the description length of shape coordinates, in order to present (see the the next section) a conservative estimation of the precision range in which the Decoupled coding prevails.

\section{Results \label{results}}

\subsection{Results of the description length analysis}
\subsubsection{Information gap for known facial images in the training set, at different resolutions}

In this section, we analyse how the efficiency of the $\Rdue$ coding varies as a function of the resolution of the dataset images. We expect that the information gap increases with the resolution. If the resolution is so low that the distance between pixels (normalised to the image height, $h^{-1}$), is of the same order of the typical deviation of landmark coordinates from their average, $\<\ell_i^2\>^{1/2}$, the uniformation will not have an effect and consequently the $\Rdue$ code may not be worth in terms of description length. In the opposite situation, $h^{-1} \ll \<\ell_i^2\>^{1/2}$, we expect a larger information gap. 

\begin{figure}
\begin{center}
    \makebox[\textwidth][c]{\includegraphics[width=\textwidth]{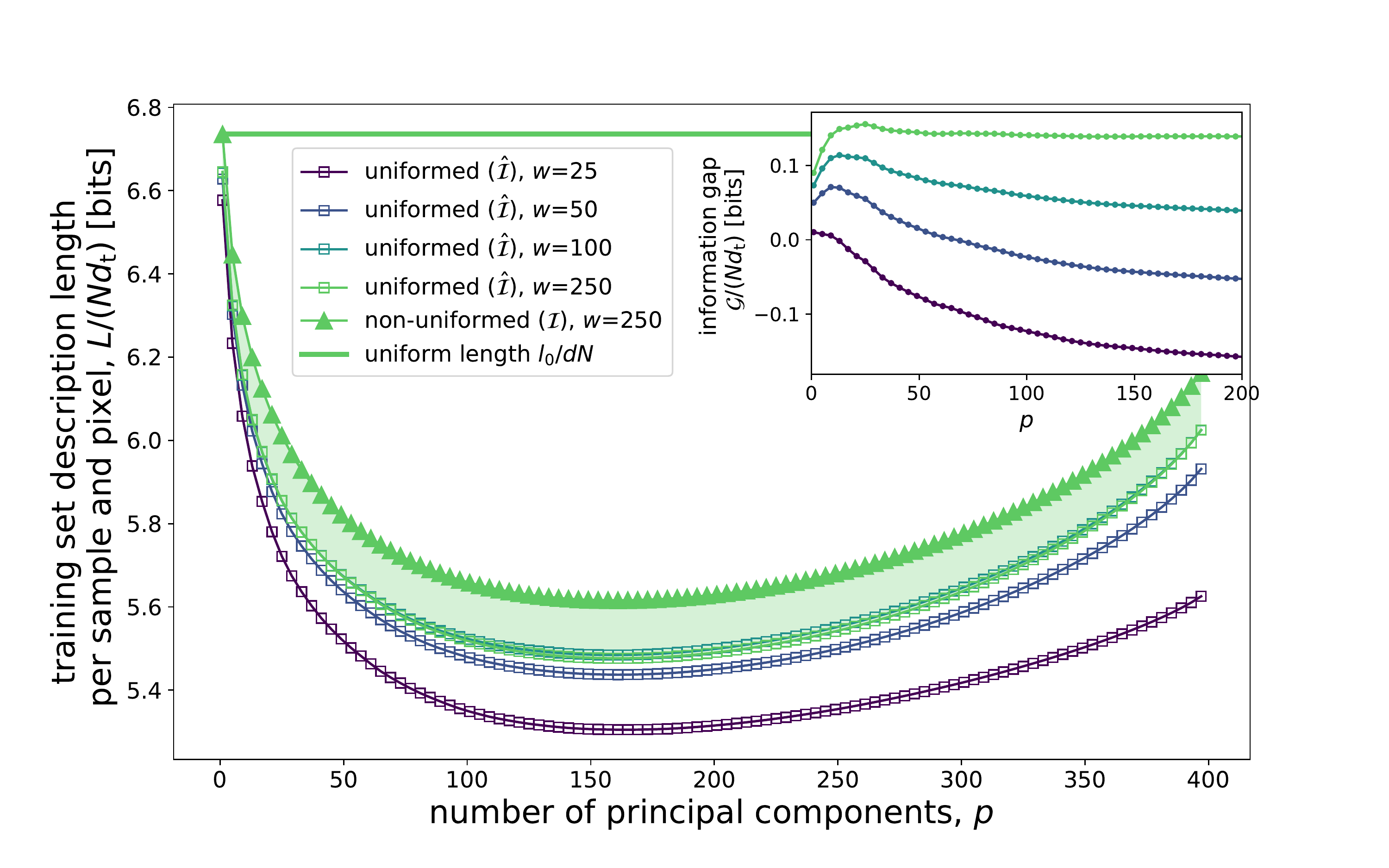}}
	\caption{Description length of uniformed images in the training set. See the main text for explanation.}
    \label{fig:trainingevidences}
\end{center}   
\end{figure}

To test this hypothesis, we calculate $p^*$ for every kind of coordinate and resolution, as the minimum of the $L_p$ curves. $p^*$ results to be lower than $N$ in the three kinds of coordinates (shape, non-uniformed texture, uniformed texture). Figure~\ref{fig:trainingevidences} shows the description length of uniformed images in the training set (i.e., taking $\hat{\cal I}=\hat{\cal I}_{\rm tr}$ in equation \ref{eq:gain}, where $\hat{\cal I}$ is the whole database of $N=400$ uniformed smiling and neural images) as a function of $p$, and for four different resolutions. 

The description length of the image dataset is slightly over-linear in the number of pixels $d_{\rm t}$, as shown by the lack of superposition of the curves in figure \ref{fig:trainingevidences}. Indeed, the largest images actually contain more information per pixel: this is the information that, according to the $p$-PCA model, has been lost when lowering their resolution to construct the lower-resolution datasets. 

As a reference value for the analysis of description length curves, it is useful to compare the values in the figure with the {\it uniform length} $l_0/(d_{\rm t}N)$, or the minimum amount of information per sample and pixel that would take to store a dataset consisting in images whose pixels fluctuate {\it independenly} around their average value in an interval of length $R$, being $R$ such that the variance per pixel is equal to the empirical average variance $\bar v_{\rm t}$ of the dataset $\cal I$ (roughly equal to $37$ units per pixel out of $256$ in $8$-bit grayscale encoding).\footnote{In other words, if one assumes that the pixel values are uniformly distributed around their average in a $d_{\rm t}$-dimensional hypercube of size $R=(12 \bar v_{\rm t})^{1/2}$, then $l_0/dN=(1/2)\log_2(12 \bar v_{\rm t})-\log_2 \epsilon$. This value is very close to the empirical entropy of the dataset corresponding to a PCA model with $p=0$ (see the Supporting information): $L_0={\sf S}_0=(1/2)\{\log_2(2\pi)+1+\log_2(\tilde v)\}-\log_2 \epsilon$, where $\log_2 \tilde v=\overline{\log_2 \lambda}$ (see the proximity of $L_0$ and $l_0$ in figure \ref{fig:trainingevidences}). } 

Figure \ref{fig:trainingevidences} also shows the description length of non-uniformed images for the largest resolution, $w_{\rm max}\times h_{\rm max}=250\times 300$. We see that, for this resolution, the information gap in equation \ref{eq:gain} is positive: the uniformed images are better compressed than non-uniformed images, for all values of $p$. The information gap per sample and coordinate ${\cal G}/(Nd_{{\rm t}})$ is, as expected, an increasing function of the resolution -- see the inset of Figure \ref{fig:trainingevidences} -- indicating that the information gap increases faster than linear in $d_{\rm t}$. Rather, for the two lowest resolutions, decoupled coding does not lead to a gain in information. Indeed, for $w=25$ the information gap is negative, roughly equal to minus one hundred of bits per sample.

\begin{figure}
\begin{center}
    \makebox[\textwidth][c]{\includegraphics[width=\textwidth]{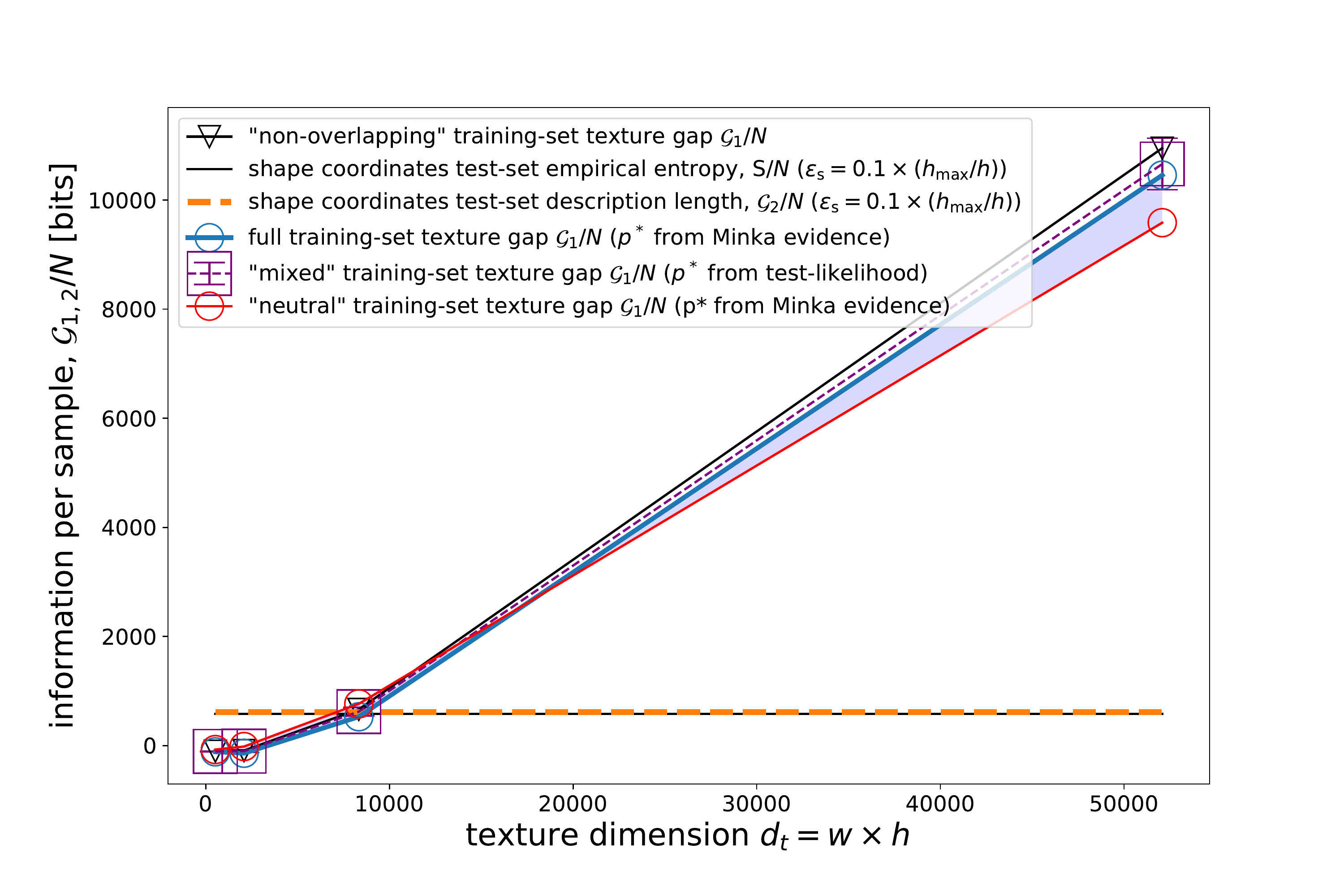}}
	\caption{Information gap for facial images in the training set. See the main text for details.}
    \label{fig:gap_info_vs_resolution}
\end{center}   
\end{figure}

The information gap per sample of the $\Rdue$ coding increases rapidly with the number of pixels $d_{\rm t}$, and it reaches more than $10000$ bits per image for $w=250$. This is evident in Figure \ref{fig:gap_info_vs_resolution}, which shows the information gap per sample ${\cal G}/N_{\rm tr}$ of training set images as a function of the resolution.  Figure~\ref{fig:gap_info_vs_resolution} also shows the description length of shape coordinates, $L_{{\cal L}_{\rm tr},p^*_{\rm s}}({\cal L}_{\rm tr})$, which is independent of the image resolution (horizontal line, see the details in the Supporting information). The information gap of the image degrees of freedom results comparable with the shape coordinates' description length $L_{{\cal L}_{\rm tr},p^*_{\rm s}}({\cal L}_{\rm tr})$ for $w=150$, but it is much larger for the largest resolution $w=250$.  Summarising, for the largest image resolution, the texture-shape decoding entails a gain in information for the largest resolution. For $d_{\rm t}\gtrsim 10^4$, the condition in equation \ref{eq:conditionongain} is satisfied.

A significant observation is that the standard deviation per pixel, $\bar v_{\rm t}^{1/2}$, of the order of $37$ units, is {\it not significantly smaller} in the dataset of uniformed images $\hat{\cal I}$. This means that uniformed images are more easily compressible, not simply because the dataset is less-varying, or more homogeneous, but because of the presence of stronger pairwise correlations between pixels in the uniformed images. Stronger correlations induce a more inhomogeneous spectrum of $C^{({\rm t})}$, say $\lambda^{({\rm t})}_1,\ldots,\lambda^{({\rm t})}_p$ and, consequently, a lower empirical entropy $\sf S$ of the associated Gaussian distribution, which, up to an additive constant, depends on the eigenvalues (see the Supplementary information) as $(1/2)\sum_{i\le p}\ln\lambda^{({\rm t})}_i$. Indeed, for the highest resolution, the excess of standard deviation per pixel and sample of non-uniformed images is $\simeq 0.8$. Neglecting the correlations, this would amount to an increment of {\it uniform length} $l_0/(d_{\rm t}N)$ (or, equivalently, of $L_0/(d_{\rm t}N)$) of $\simeq 0.04$, which is one order of magnitud lower than the texture gap ${\cal G}_1$.

As a consistency analysis, we have calculated the information gap in two different ways: first, taking $p^*$ according to Bayesian model selection, $p^*=\arg\max_p L_{{\cal D}_{\rm tr},p}({\cal D}_{\rm tr})$; second, taking the value that maximises the validation-set (out-of-sample) likelihood, by $K$-fold cross-validation of the validation/training separation of the original dataset (c.f. details in the Appendix). Both ways of computing the training-set information gap are consistent within the (cross-validation) statistical errors (which is not obvious, specially considered that $d_{\rm t}\gg N$). 

\subsubsection{Information gap for known facial images in the training set that show different facial expressions} 

In this analysis, we test the hypothesis proposed in the introduction that decoupled coding is particularly effective when encoding variants of known facial images which differ only in facial expressions. By definition, variations in facial expression are expected to change mainly shape coordinates, and much less texture coordinates (that are independent of the positions of the landmarks and hence nearly independent of the facial expression). The information gap should increase in this situation, since the texture coordinates of facial images differing in expression should be more redundant, correlated and easily compressed -- or, in the language of probability, they should exhibit a larger likelihood.

To test this hypothesis, we computed the training-set information gap for two dataset of length $N_{\rm tr}=200$: the former (called ``neutral'') consisting of neutral expression images of 200 different subjects and the latter (called ``mixed'') corresponding to both the neutral and the smiling portraits of the same $100$ (randomly selected) subjects. The blue shadowed area in figure \ref{fig:gap_info_vs_resolution} indicates the difference in information gap between the ``mixed'' and the ``neutral'' training sets. While the information gap of the ``mixed'' dataset is indistinguishable from the ``full'' dataset of $N_{\rm}=400$ images, the ``neutral'' dataset presents a lower information gap. 

This analysis is consistent with our initial hypothesis. Notice that this result is not a trivial consequence of the fact that the ``mixed'' dataset (consisting in $N=200$ portraits of {\it the same $100$ subjects}) is more easily compressible than the ``neutral'' one (consisting in $N=200$ portraits of $200$ {\it different subjects}): indeed, for {\it both uniformed and non-uniformed} facial images, the description length of the ``mixed'' dataset is lower than that of the ``neutral'' dataset. What is less trivial is that {\it the gap $\cal G$ is higher for mixed images}. 

\subsubsection{Information gap for unknown facial images in the test set that show different facial expressions}
\label{sec:info_gap_novel}

Here, we perform a variant of the above analysis aimed to test that the decoupling is particularly efficient when encoding {\it unknown} (not belonging to the training set) facial images that correspond to subjects that {\it are present} in the training set, with a different facial expression. 

We have already seen that the training-set of uniformed images exhibits lower description length (and empirical entropy) than the training set of raw dataset images. It is hence reasonable to suppose that decoding does not only reduce the {\it bias error} (of the training-set) but also the {\it variance error} in the description of unknown facial images, belonging to a test-set.\footnote{In the language of probability, we have seen before that the uniformed images present stronger between-pixel correlations $C_{ij}$ while presenting a roughly equal total variance (or $\traza(C)$). This is the reason for which, for uniformed faces, the training-set empirical entropy ($\sum_i \ln\lambda_i$, up to a constant) is lower (hence the likelihood is higher). A lower test-set entropy would simply imply that also the term $\traza(C_{\rm te}\cdot C_{\rm tr}^{-1}/2)$ (the difference between test- and training-set entropies, up to a constant) is significantly lower. We will call bias and variance terms of the entropy to the terms $\sum_i \ln\lambda_i$ and $\traza(C_{\rm te}\cdot C_{\rm tr}^{-1}/2)$, respectively.\label{footnote:biasvariance}} 

To test this hypothesis, we calculated the information gap in a test-set $\cal I$ in Equation \ref{eq:gain}, which is composed by $N/K=80$ (with $K=5$) images corresponding to smiling subjects, {\it whose neutral-expression images do belong to the training-set}. Notice that we will call such a set simply ``test-set''. All the information-theoretical quantities are then cross-validated for different $K$ training/test partitions of the original dataset (by means of the $K$-fold algorithm of cross-validation).

\begin{figure}
\begin{center}
    \makebox[\textwidth][c]{\includegraphics[width=\textwidth]{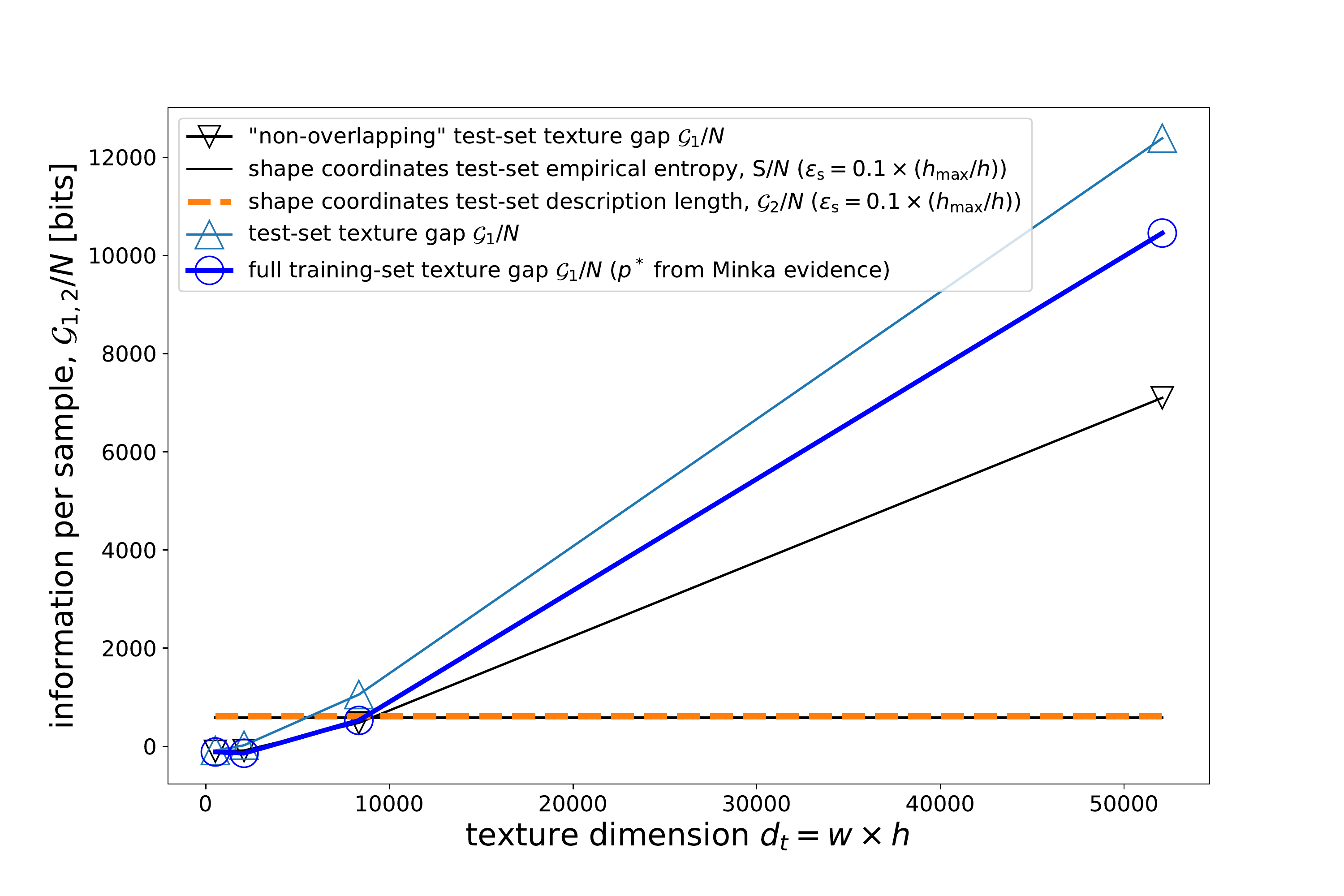}}
	\caption{Information gap for facial images in the test set. See the main text for details.}
    \label{fig:gap_info_vs_resolution_test}
\end{center}   
\end{figure}

Figure \ref{fig:gap_info_vs_resolution_test} reveals that the information gap of the test-set is significantly higher compared to the information gap of the training-set, with a p-value lower than $10^{-4}$ (notice the small errorbars of the training-set information gap in figure \ref{fig:gap_info_vs_resolution}). The increment in information gap per sample (roughly 1/6 of the test-set gap) corresponds to the bits that one saves using the decoupled coding for unknown smiling faces, not belonging to the training-set. This implies, as expected, that the $\Rdue$ coding reduces both the bias and the variance errors for variants of known faces differing in facial expression only. {\it $\Rdue$ is, hence, particularly efficient}, in terms of information, {\it to encode facial images of known subjects, differing in facial expressions}. 

It is interesting to compare this result with the texture information gap of a different test-set, which we call ``non-overlapping'', in which the single folds are composed by $N/K=80$ facial images corresponding to $40$ subjects with  both smiling and non-smiling expression. The non-overlapping test-set contains, in this way, images of subjects {\it that are not present in the training-set} (so that test- and training-sets contain information regarding different subject identities). Figure \ref{fig:gap_info_vs_resolution_test} shows how the texture information gap ${\cal G}_1$ of the non-overlapping dataset is even lower than the training-set gaps. This result shows that the decoupling code $\Rdue$ {\it is less efficient to encode unknown facial images corresponding to unknown subjects}. Furthermore, this results shows that while uniforming variants of known faces that differ only in facial expression implies a reduction of both the bias and the variance terms of the entropy (see footnote \ref{footnote:biasvariance}), uniforming facial images of unknown individuals leads to a reduction of the bias term {\it but to a positive increment of the variance term}. In any case, we stress that 
the texture information gap ${\cal G}_1$ is still larger than ${\cal G}_2$ for the non-overlapping test-set. Consequently, the decoupling code is more efficient even when processing unknown-identity facial images, albeit in this case the description length gap is lower.

\subsubsection{Summary of the results of the description length analyses}

In sum, our analysis shows that decoupled coding leads to a more efficient encoding of known facial images (i.e., in the training set) compared to eigenface coding, when the images are shown at high enough resolutions and in particular when they differ in facial expressions. Furthermore, our results show that the efficiency of the decoupled coding is magnified when the task consists in encoding {\it unknown} variants of known faces differing in facial expression.

\subsection{Analysis of the performance of decoupled and eigenface coding in face processing tasks}\label{sec:classification_experiments}

So far, we have used the normative construct of description length to assess the efficiency of decoupled coding. Here, we ask how the normative advantage of decoupled coding translates into a better performance in facial processing tasks and what are exactly the advantages. For this, we compare the performance of eigenface and decoupled coding in three face processing simulations that help illustrate the most important differences between the coding schemes; namely, (1) sampling artificial facial images from the learned generative model, (2) recognizing facial identity, and (3) reconstructing unknown faces. Please see the Supporting information for a supplementary (gender classification) simulation. 

\subsubsection{Simulation 1: Sampling synthetic faces from the trained generative model}

The generation of artificial faces is a widely used task in AI to demonstrate the quality of a learning algorithm or encoder. In this simulation, our goal is not to challenge the performance of mainstream machine learning approaches that use deep nets with millions of parameters \cite{dosovitskiy2016,otoole2018,suchow2018learned,liu2019}, but rather to test the hypothesis that a very simple (20 degrees of freedom) linear model can generate realistic images, when it is based on decoupled coding.

Each PCA-based representation of the training set ${\cal I}$ induces a simple generative model of faces (see the Supporting information for details). In particular, $\Runo$ induces a multivariate Gaussian distribution in the space of facial images. Rather, $\Rdue$, induces two separate Gaussian distributions over uniformed texture and shape coordinates, respectively. It is possible to create {\it synthetic} facial images by sampling from the respective probability distributions of $\Runo$ and $\Rdue$. In the case of $\Rdue$, after sampling from both probability distributions, it is necessary to de-uniformize the sampled uniformed texture coordinates given the sampled shape coordinates (see the Supporting information for details about the de-uniformation procedure). 

\begin{figure}
\begin{center}
    \makebox[\textwidth][c]{\includegraphics[width=\textwidth]{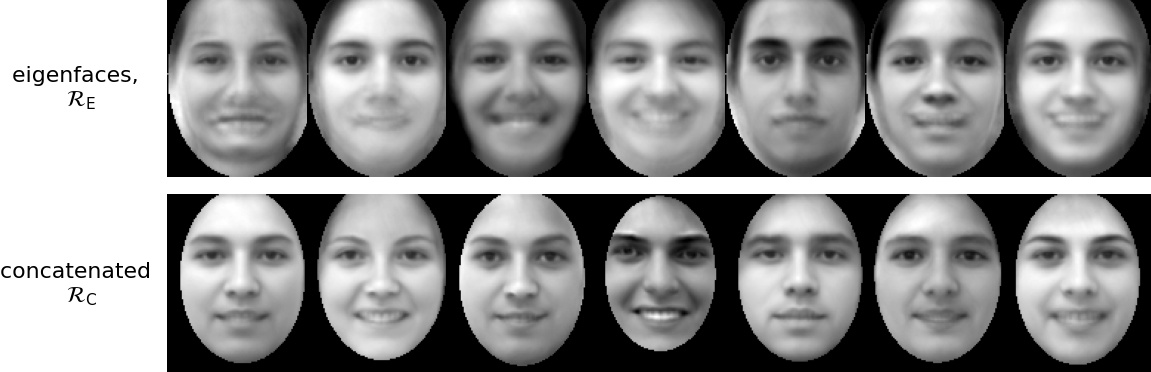}}
	\caption{Examples of synthetic faces sampled from the generative models of $\Runo$ (top row) and a simple variant of the decoupling $\Rdue$ coding scheme, which we call concatenated coding, $\Rtre$ (bottom row). See the main text for details.}
    \label{fig:sampling}
\end{center}   
\end{figure}

Figure \ref{fig:sampling} shows example synthetic images created by sampling $\I$ and $(\hat{\I},{\bm \ell})$ from the models induced by $\Runo$ and $\Rdue$, respectively. In both cases, we used $p=20$ degrees of freedom, randomly chosen among the first $40$ principal components of each model.\footnote{In particular, we sample $20$ principal components $x'_i$ from their respective distributions, where the index $i$ may take the values $1,\ldots,p=40$. The remaining $20$ coordinates among the first $40$ coordinates are set to zero.}  Please note that, the larger the value of $p$, the higher the dimension of (the vector space of) the sampled facial images. When small values of $p$ are used, the generative models produce low-dimensional variations of the average face; this implies that the synthetic faces are realistic (free from artefacts) but very stereotyped, with low variability. Rather, using larger values of $p$ is a more compelling task, since the generative models are free to produce faces with high variability -- but at the same time it is harder for them to produce realistic faces that are free from artefacts.

Figure \ref{fig:sampling} permits appreciating that with a relatively high value $p=20$, both eigenface coding ($\Runo$) and decoupled coding ($\Rdue$) produce produce faces with high variability. However, only the faces produced by the latter are realistic and free from artefacts. This simulation therefore shows that a very simple linear model based on decoupled coding (but not on eigenface coding) can produce realistic and varied facial images.

Please note that for this comparison we consider a simple variant of the decoupling $\Rdue$ coding scheme, which we call {\it concatenated coding}, $\Rtre$, consisting in the principal components of the concatenated set of texture and shape coordinates (see the Supporting information in section~\ref{sec:supplementary} for a formal description). The reason is that $\Rtre$ permits choosing a single number of principal components, $p$, in common with $\Runo$ -- which therefore permits comparing the two codes with the same number of parameters. The use of $\Rdue$ would require, instead, to choose separately $p_{\rm t},p_{\rm s}$ for a fixed $p_{\rm t}+p_{\rm s}=p$. Indeed, the results of figure \ref{fig:sampling} are qualitatively identical if one directly compares $\Runo$ with $\Rdue$, varying $p_{\rm t}=p$ and fixing $p_{\rm s}$ to its maximum value, $=d_{\rm s}=92$). We remark that the concatenated coding is considered here and in the next subsection only for the sake of simplicity of the $\Runo$-$\Rdue$ comparison\footnote{Since $d_{\rm t}\gg d_{\rm s}$, and since, as we mention in the Supplementary information, both coordinates are weakly correlated, the eigenvalues of $C^{({\rm c})}$ are dominated by those of $C^{({\rm t})}$ and the likelihoods almost coincide. For this reason, and because of the lack of a biological motivation, we do not present to present an information-theoretical analysis of the concatenated model.}.

\subsubsection{Simulation 2: Facial identity recognition}\label{sec:facial_identity}

The results of the description length analysis show that decoupled coding may be particularly suited to encode efficiently natural variants of known facial images, which consist in variations of facial expressions. This is because variants of a known face may affect one set of coordinates shape or texture, while leaving the other essentially unaffected. For example, recognising a known face with a different facial expression might benefit from the reuse of uniformed texture coordinates, which would be relatively unaffected. 

In this simulation, we test whether and how the description length advantage of decoupled coding translates into a better capability to recognise faces with different facial expressions. For this, we exploit the fact that the FEI database includes $400$ facial images of $200$ subjects, with 2 images of the each subject that only vary for facial expressions: neutral or smiling. The simulative task consists in recognising which of the $200$ images showing a neutral expression corresponds to a target image (excluded from the training set) in which the same person shows a smiling expression, and vice-versa. 

We implement the recognition task through a nearest-neighbour classifier, using a {\it distance} ${\sf d}_p({\bf z},\x(s))$ among the facial coordinates of the target smiling face $\bf z$, and those of all the $200$ neutral expression images $\x(s)$. The distance ${\sf d}_p(\cdot,\cdot)$ among facial coordinates is the Mahalanobis ($+L_2$) distance, known to be a better measure of facial similarity than the simple Euclidean metric between images \cite{Moghaddam1998}\cite[Chs.~5-6]{Wechsler2007}\footnote{Although surely not the most efficient method for a supervised classification analysis, we choose the Mahalanobis distance algorithm, since it is the one that uses the only the information defining our working models, i.e., $C_p$ for each kind of coordinate.}. The Mahalanobis metrics between a couple of vectors is a scalar product between their (normalised) first $p$ principal components,\footnote{\label{footnote_distance} ${\sf d}_p({\bf u},{\bf v})=[({\bf u}-{\bf v})^\dag\cdot C_p^{-1} \cdot ({\bf u}-{\bf v})]^{1/2}$, where $C_p=E^\dag_p\cdot \Lambda_p\cdot E_p$, and where $\Lambda_p$ is the diagonal matrix of the largest $p$ eigenvalues.}, which does depend on the eigenvalues and eigenvectors of the correlation matrix and, hence, on the training set. It is important to remark that that the target image coordinates are excluded from the training set (see the details in the Supporting information). 

\begin{figure}
\begin{center}
    \includegraphics[width=0.85\textwidth]{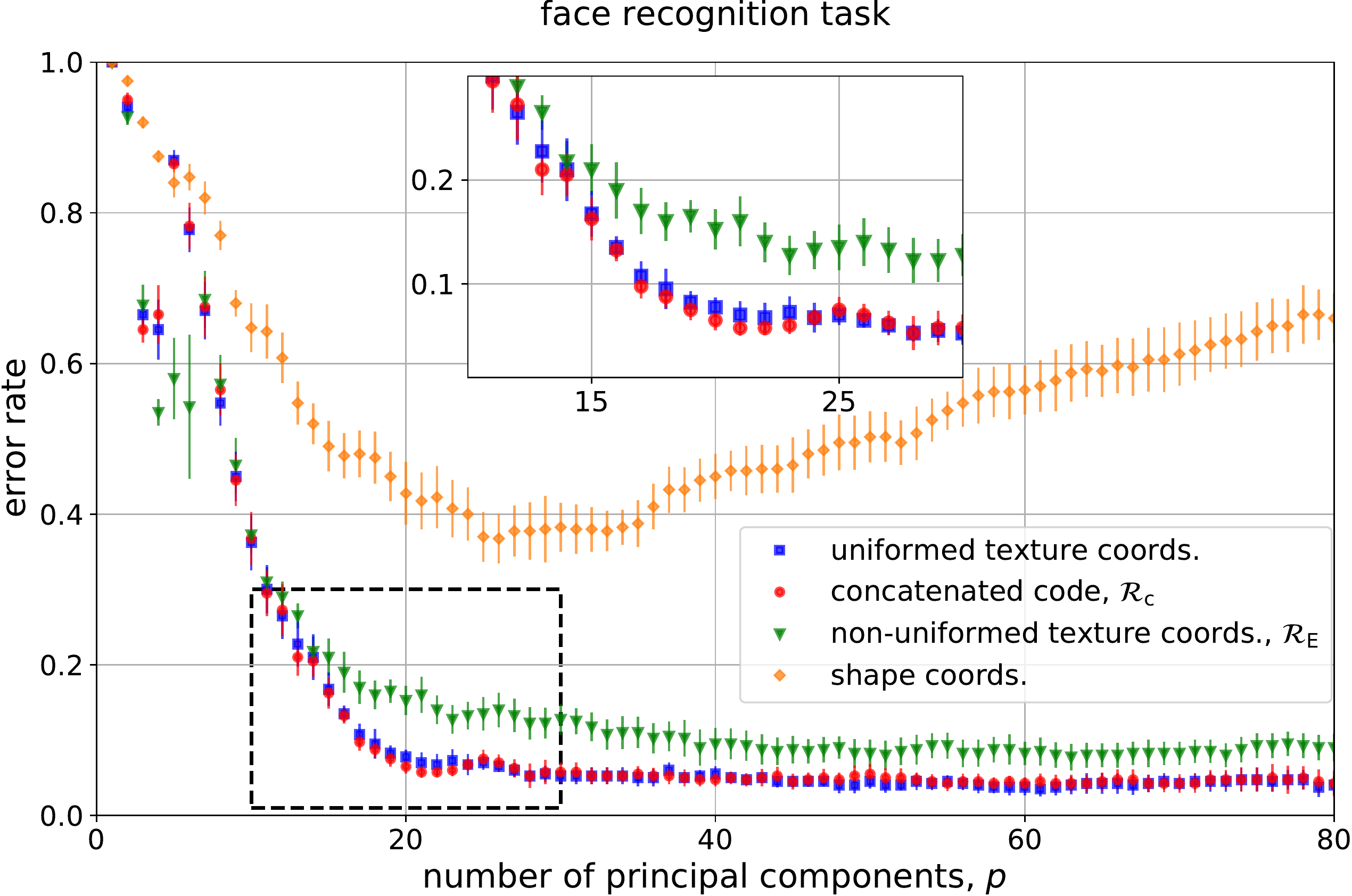}
    \caption{Performance of in the face recognition task using the uniformed images, the non-uniformed images, and the shape coordinates. See the main text for explanation.}
    \label{fig:recognition}
\end{center}   
\end{figure}

Figure \ref{fig:recognition} shows the results of the facial identity recognition task. The figure reports the misclassification error as a function of the number $p$ of principal components considered, for each kind of coordinate (uniformed images, non-uniformed images and shape coordinates, or $\x=\I$, $\hat\I$, ${\bm \ell}$, respectively). The errors decrease and reach a plateau in the cases of both non-uniformed and uniformed images, but the latter consistently exhibits a better performance. Interestingly, this simulation permits appreciating the relative contributions of texture and shape coordinates to the task. As expected, the performance of shape coordinates is significantly lower, with a minimum error around $0.4$ (notice that in such recognition task, the random choice error rate is $1-1/N_{\rm train}\simeq 0.996$, see the Supporting information). However, we notice that this fact depends on the arbitrary choice of the image resolution (determining the image dimension $d_{{\rm t}}=w\times h$) and on the number of landmarks $\nl$  (determining the shape coordinate dimension $d_{\rm s}=2\nl$). Using a larger number of landmarks will enhance the relative relevance of shape coordinates.
Moreover, figure \ref{fig:recognition} reveals as well that the distance based on the concatenated code $\Rtre$, which exploits the correlation between shape and uniformed texture coordinates, does not perform significantly better than that based on uniformed images. This is due to the fact that the images contain more information than the shape coordinates (since $d_{\rm t}\gg d_{\rm s}$), and that shape and texture coordinates are only weakly correlated (see the Supporting information). 

The reader may find a discussion on the qualitative differences in the shape of the error rate curves of texture and shape coordinates in the Supplementary information.


\subsubsection{Simulation 3: Reconstructing novel faces}

The reconstruction task consists in representing novel facial images that do not belong to the training set, in terms of an expansion in $p$ principal components only. In mathematical terms, if $\x$ is a facial image, or its shape coordinates, the reconstruction of $\x$ in terms of the first $p$ principal axes is $\x_p=E_p^\dag\cdot E_p\cdot \x$, where $E_p$ is the $p\times d$ matrix of the first $p$ (row) eigenvectors.\footnote{If $\x$ corresponds to an image, the reconstructed image is different from the original one even with $p=N$ coordinates, since the matrix $E_p^\dag\cdot E_p$ is different from the identity matrix, as far as it has rank $=p\le N< d_{{\rm t}}$.} In the case of the $\Runo$ code, we simply project the original image ($\x=\I$) onto the first $p$ eigenfaces, $\x_p=E_p^\dag\cdot E_p\cdot \x$. In the case of $\Rtre$, we perform this operation for both $\hat\I_{p_{\rm s}}$ and ${\bm\ell}_{p_{\rm t}}$ and, afterwards, we perform a de-uniformation (see the Suplementary material), leading to a non-uniformed reconstructed image  $(\hat\I_{p_{\rm s}},{\bm\ell}_{p_{\rm t}}) \to \I$.

\begin{figure}
\begin{center}
    \makebox[\textwidth][c]{\includegraphics[width=\textwidth]{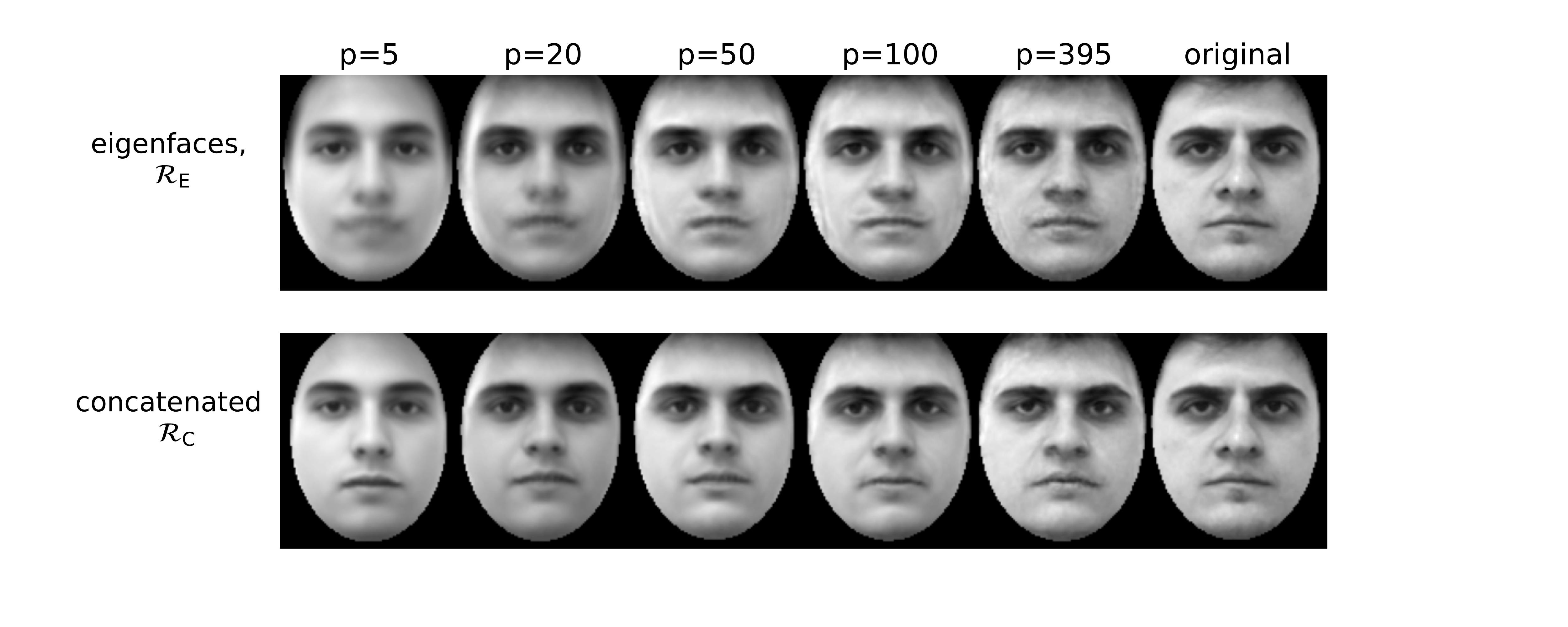}}
	\caption{Reconstruction of a test-set facial image with $p$ principal components according to the representations: eigenface coding ($\Runo$, first row) and concatenated coding ($\Rtre$, second row). The 5 columns of the matrix of images represent, respectively: $p=5$, $20$, $50$, $100$, $395$, and the original image. The image is $100\times 120$.}
    \label{fig:reconstruction}
\end{center}   
\end{figure}

Figure \ref{fig:reconstruction} shows the reconstruction of a novel face according to $\Runo$ and $\Rtre$ for different values of $p$. The figure illustrates that $\Runo$ produces a border artifact; this is because linear combinations of different facial images with different facial contours result in an image which tend to be blurred in the margin of the face. Rather, the border artifact is absent for $\Rtre$, and the representation for high $p$ is slightly more faithful. 

\begin{figure}
\begin{center}
    \makebox[\textwidth][c]{\includegraphics[width=0.75\textwidth]{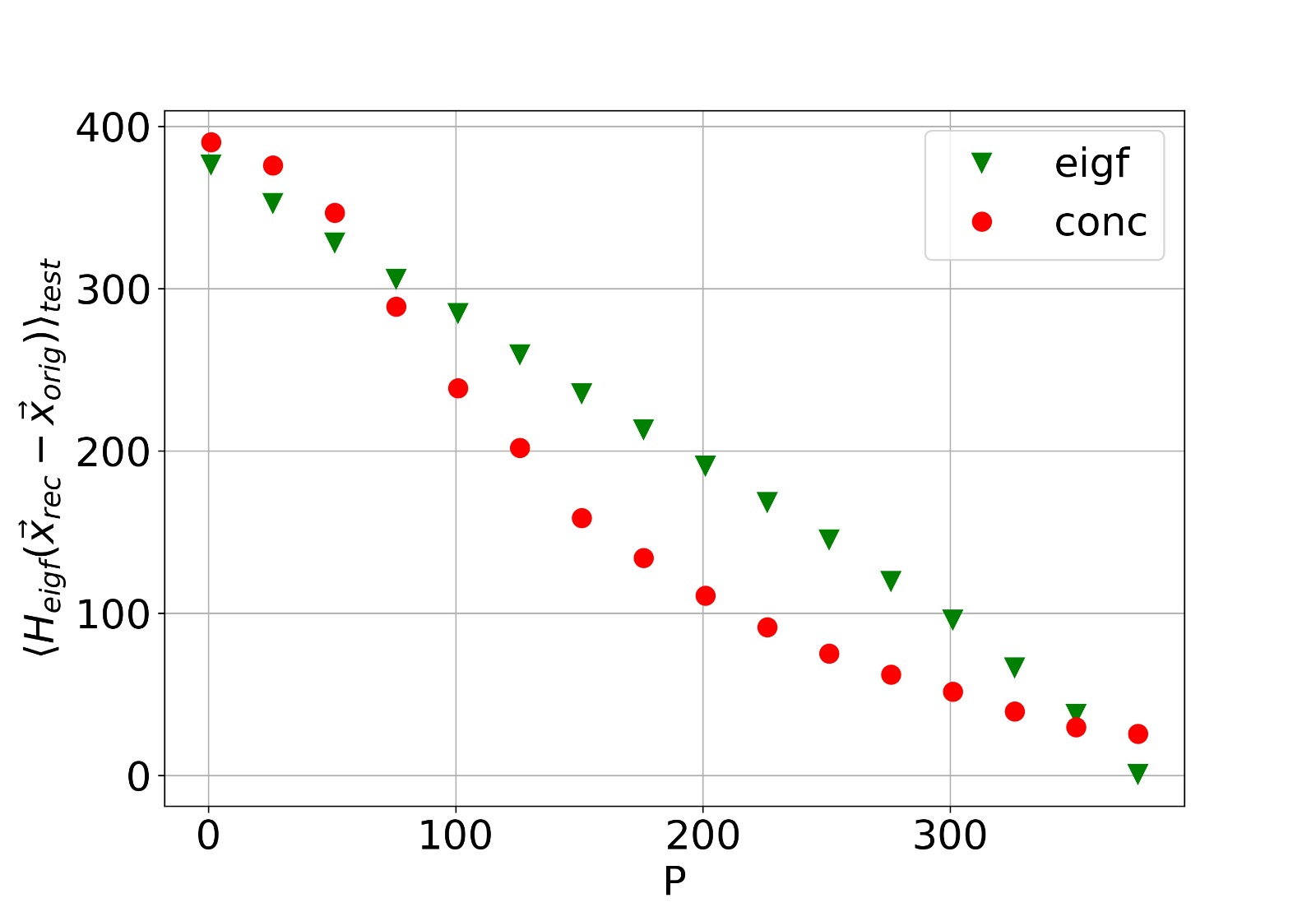}}
	\caption{Mahalanobis distance ${\sf d}_{C,p_d}({\bf t}_p,{\bf t})$ among target facial images and their reconstructed counterparts with $p$ principal components (in the abscissa), according to the codes $\Runo$, $\Rtre$ (see the main text for details). The points are the average of the distance over $N_{\rm te}=20$ randomly selected images with neutral facial expression.}
    \label{fig:reconstruction_distance}
\end{center}   
\end{figure}

This is quantitatively illustrated in figure \ref{fig:reconstruction_distance}, which reports the Mahalanobis distance (see footnote \ref{footnote_distance}) between the target facial image, $\bf t$, and its reconstruction from $p$ principal components, ${\bf t}_p$, according to the representations $\Runo$ and $\Rtre$. Crucially, in this analysis the target vectors are $N_{\rm te}=20$ neutral, randomly chosen test-set images, {\it not belonging to the training set} from which the ``reconstructing matrix'' $E_p^\dag\cdot E_p$ are calculated. Also crucially, the matrix $C$ defining the Mahalanobis distance ${\sf d}_{C,p_{\rm d}}({\bf t}_p,{\bf t})=H_{C,p_{\rm d}}({\bf t}_p-{\bf t})$ with $H_{C,p_{\rm d}}(\x)=\x^\dag\cdot {C_{p_{\rm d}}}^{-1}\cdot \x$ is taken {\it as the (training-set) correlation matrix of the eigenface coding}, with the maximum number of principal components $p_{\rm d}=N_{\rm tr}-N_{\rm te}-2=378$. Indeed, for a numerical assessment of the similarity between reconstructed and original faces, one needs a perceptual distance, based on a dimensionality reduction, hence, based on a representation. In fact, we have observed that, as expected \cite{valentine2016}, the simple Euclidean metrics, or the pixel-wise distance, does not allow to assess any difference between the two representations. Choosing the {\it representation used to compute the similarity between target and reconstructed images equal to one of the representations used for the reconstruction} (the eigenface code $\Runo$), we are biasing the comparison in favour of this code. Despite the unfavorable bias towards the $\Rtre$ code, it exhibits a lower perceptual distance in figure \ref{fig:reconstruction_distance} for all the values of $p$ except for the maximum $p=p_{\rm d}$ {\it for which, by construction, the $\Runo$ code presents the minumum possible distance ${\sf d}=0$} (since the reconstruction and the target image are expressed in terms of the same principal axes){\it , and hence it is not possible that a different model improves it.} It is remarkable that, as far as $p<p_{\rm d}$, the $\Rtre$ code improves $\Runo$ despite such a bias.

As for Simulation 1, the results of this subsection are qualitatively identical if one directly compares $\Runo$ with $\Rdue$, varying $p_{\rm t}=p$ and fixing $p_{\rm s}$ to its maximum value, $=d_{\rm s}=92$).

\subsubsection{Summary of the results of the face processing tasks}

In sum, our analysis shows that a tiny model using decoupled coding and only 20 degrees of freedom can be sampled to produce realistic synthetic images, whereas sampling from a model using eigenface coding produces less realistic faces with artefacts. Furthermore, decoupled coding greatly facilitates the recognition of familiar faces with novel facial expressions -- especially thanks the fact that texture coordinates remain stable across different expressions. Finally, decoupled coding outperforms eigenface coding in the reconstruction of novel faces. Please see the Supporting information for an additional simulation of gender classification using the two coding schemes.


\section{Discussion and Conclusions}

\label{sec:discussion}

Recent research in neuroscience claims that the neural coding for facial identity in the inferotemporal (IT) cortex of macaques \cite{chang2017,chang2021explaining} implements a decoupled coding scheme in which distinct subpopulations of neurons project facial images onto two distinct sets of axes, which encode the geometric shape of a face and its texture separately. From a computational perspective, decoupled coding affords accurate face processing, permitting the linear decoding of facial features from single cell responses; and it outperforms widely used schemes in vision research, such as eigenface coding \cite{chang2017,chang2021explaining,higgins2021}.

In this article, we aimed to elucidate the normative reasons for this advantage, by appealing to the notion of \emph{description length}, which permits quantifying the efficiency of neural coding schemes in info-theoretic terms. The general idea is that the best model is the one that minimizes the amount of memory (bits) required to encode both the data and the model parameters themselves \cite{rissanen1978modeling,rissanen1999hypothesis}. 

In particular, we analysed the description length of the decoupled coding ($\Rdue$), aka the Active Appearance Model (AAM), which encodes both the principal components of {\it uniformed} (shape-free) facial images and their shape (landmark) coordinates. We compared such description length with that of the set of principal components of the (non-uniformed) raw images, the widely used eigenface coding ($\Runo$). At difference with previous studies that compared alternative (biologically plausible) neural codes \cite{chang2017,chang2021explaining}, here we performed an information-theoretic analysis: we evaluated the gain in information entailed by the uniformation of facial images, and compared it with the amount of information required to encode the landmark coordinates. This evaluation necessarily requires calculating the information length of (the principal components of) non-uniformed images, which is precisely what eigenface coding ($\Runo$) does -- and hence it results in an implicit comparison of decoupled coding ($\Rdue$) and eigenface coding ($\Runo$).  

Our simulations, on the same database (FEI, see \cite{fei}) as in the monkey study of \cite{chang2017}, show that decoupled coding ($\Rdue$) requires less information to represent the images compared to eigenface coding ($\Runo$), despite the latter does not require coding for the geometric coordinates of faces. Remarkably, the efficiency gain of decoupled coding ($\Rdue$) is especially prominent for high resolution images and for variants of training set images that only differ in facial expressions. 

The number of landmarks $\nl$ or, equivalently, the dimension of the shape coordinates ($d_{\rm s}=2\nl$) is, by construction, much lower than the image resolution $d_{\rm t}$. In information theoretical terms, it is precisely this condition $d_{\rm s}\ll d_{\rm t}$ that makes the decoupled coding efficient: encoding the position of few landmarks costs an information which is much lower than the information that is saved encoding the resulting {\it uniformed} facial images (for high enough image resolution). Indeed, a control analysis shows that this advantage is in place up until the landmarks become too numerous, i.e., of the order of a few hundreds (see the Supplementary information, paragraph “How many landmarks are too many?”).

Furthermore, we found that the probabilistic generative model induced by decoupled coding ($\Rdue$) achieves good performance in face processing tasks, including sampling artificial or novel faces, recognising face identity and reconstructing novel faces with $p$ principal components. Rather, a model using eigenface coding ($\Runo$) performs less accurately and produces less realistic faces with artefacts. 

Taken together, these results shed light on the normative advantages of the decoupled coding for faces that was empirically reported in monkey inferotemporal (IT) cortex of macaques \cite{chang2017,chang2021explaining}, showing that it is both more efficient (in info-theoretic terms) and more accurate than the alternative eigenface scheme widely used in computer vision. 

The dataset of uniform facial images does {\it not} present, as one may have expected, significantly lower variance than the set of non-uniform images. The efficiency of the decoupled coding comes, instead, from the fact that pixels of uniform images are more correlated and, consequently, they can be compressed in terms of few principal components without loss of precision. The amount of information saved in encoding uniform images compensates, for high enough resolution, the amount of information needed to codify the principal components of shape coordinates.

{\it Shape} (texture free) and {\it texture} (shape-free) coordinates carry information regarding naturally different aspects of human faces and can vary independently. Namely, variations in facial expressions and variations in perspective (e.g., small rotations) mainly modify shape coordinates, but not texture coordinates. Rather, variations in luminosity, suntan, or make-up only modify texture coordinates \cite{laurentini2014}. This perspective helps explain our finding that decoupled coding  is particularly advantageous when encoding {\it variants of known faces} or unknown faces (in the test set). When encoding variants of known images, one of the two sets of (shape and texture) coordinates will tend to remain the same as in the known, reference image.

Of note, the above described advantages of decoupled coding come at the price of performing some prerequisite nonlinear computations over facial images: at least landmark detection and image deformation (see the Supporting information). We speculate that if the neural code for facial identification implements a variant of the AAM \cite{chang2017}, these nonlinear computations might be putatively realised by early visual areas, which lie below the IT in the neural hierarchy. This speculation remains to be tested in future research, since the neuronal mechanisms of landmark detection and image deformation have not been identified yet. Finally, future research could address the conditions under which the decoupling emerges spontaneously in deep architectures, as well as a study of the AMM efficiency in which the cost of landmark detection and uniforming is taken into account.



\section{Supporting information}
\label{sec:supplementary}

\paragraph*{Relation with Bayesian Model Selection.} Bayesian model selection consists in choosing the model $\cal M$ that maximises the Bayesian evidence of a given dataset $\D$. The best model $\cal M$ is, equivalently \cite{mackay2003}, the one that minimises the description length $\min_{\cal M} L_{\cal M}(\D)$. To verify the validity of the condition (\ref{eq:conditionongain}), we have, instead, compared the description length of {\it different datasets}: $\cal I$, $\hat{\cal I}$, $\cal L$, according to {\it the same probabilistic model}, which corresponds to the multivariate normal distribution whose correlation matrix takes, respectively, the values $C_{p}$, $C^{({\rm t})}_{p_{{\rm t}}}$ and $C^{({\rm s})}_{p_{{\rm s}}}$, for the three datasets. Here, $C_p$ is the matrix whose $p$ largest eigenvalues and corresponding eigenvectors are the same of the sample eigenvalues and eigenvectors of the training-set of non-uniformed images and the remaining $d-p$ eigenvalues are set to a constant (see, for example, \cite{minka2000}), and the same for $C^{({\rm t})}_{p_{{\rm t}}}$ and $C^{({\rm s})}_{p_{{\rm s}}}$.

This is, hence, the opposite situation with respect to Bayesian model selection, in which one compares the evidence of the same dataset according to different models. It is important to remark that, in the present work, we do not aim to perform a comparison, on Bayesian grounds, between eigenface and decoupled codings, understood as probabilistic models {\it over the common dataset of original, non-uniformed facial images}. Indeed, the representation $\Rdue$ induces a probability distribution in the space of non-uniformed images $\I$, that is no longer a Gaussian distribution (even if the distributions over  $\hat \I$ and $\bm \ell$ are) since it involves the nonlinear image deformation operations, that we have completely neglected in our information-theoretical analysis. Within our working hypothesis, we neglect the {\it uniformation} ${\cal I},{\cal L} \to \hat{\cal I}$ (previous to the PCA) and {\it de-uniformation} ${\cal I}_p,{\cal L}_p \leftarrow \hat{\cal I}',{\cal L}'$ (posterior to the PCA) operations from the information-theoretical analysis. 

In other words, here we consider three probabilistic descriptions over separate spaces: non-uniformed images, uniformed images and shape coordinates. Our conclusions merely rely in the information-theoretical interpretation of the Bayesian evidence of a dataset according to a model, that is related to the amount of information needed to store the dataset in terms of the model lattent variables, with a given precision.  The decoupled code, understood as a probabilistic model {\it on the space of the original images} would be more complex than Gaussian. It would implicitly contain, in some (texture) latent variables, a description of the input image somehow invariant under shape transformations; other (shape) latent variables would be invariant under texture transformations of the original image. Our current analysis is to be understood as an estimation of the information theoretical gain of the facial representation in terms of (principal components of) uniformed images and landmarks, neglecting the non-linear operations of landmark detection and image deformation that lead to these two facial coordinates, from the original dataset of images.

Instead, we do perform a genuine Bayesian model selection when choosing the values of ${p}$, ${p_{{\rm t}}}$, ${p_{{\rm s}}}$ that minimise the description length (that maximise the Bayesian evidence) of each dataset, i.e., of each type of coordinate.

\paragraph*{Image uniformation and de-uniformation.} The creation of the {\it uniformed} texture $\hat{\I}$ coordinates from the original images ${\I}$ and their shape coordinates $\bm\ell$ in $\Rdue$ is implemented, as said before, through image deformation algorithms based on similiarity transformations \cite{image_deformation}. Such algorithms map the original image into an image whose landmark positions $\bm \ell$ will now occupy their average value in the dataset $\<{\bm\ell}\>$. Vice-versa, the reconstruction of novel images in $\Rdue$ requires creating a non-uniformed facial image from the reconstructed shape and texture coordinates ${\bm\ell}_p,\hat\I_p$. This operation we will be called {\it de-uniformation}: 

\begin{eqnarray}
({\bm \ell},\I) &\begin{subarray}{c} \<{\bm \ell}\> \\ \rightarrow \end{subarray}& (\<{\bm \ell}\>,\hat\I) \qquad \textrm{uniformation} \\
	({\bm \ell},\I) &\begin{subarray}{c} {\bm \ell} \\ \leftarrow \end{subarray}& (\<{\bm \ell}\>,\hat\I) \qquad \textrm{de-uniformation}
\end{eqnarray}
where the subscripted arrow indicates the image deformation algorithm transforming an image $({\bm \ell}_1,\I_1)\substack{{\bm\ell}_2 \\ \to} ({\bm \ell}_2,\I_2)$ so that the pixel values of $\I_2$ in the positions given by ${\bm \ell}_2$ are those of $\I_1$ in ${\bm \ell}_1$ (say, $\I_2(\vec{{\ell}_2}_j)=\I_1(\vec{{\ell}_1}_j)$ where $\vec{{\ell}_1}_j$ are the original Cartesian positions of the $j$-th landmark), and the rest of the pixel values of $\I_2$ are changed consequently, under the requirement of smoothness. As a consistency check, we have verified that uniforming and consequently de-uniforming dataset images, leads to new images that are visually indistinguishable from the initial ones. 

\begin{figure}
\begin{center} 
  \includegraphics[width=10cm]{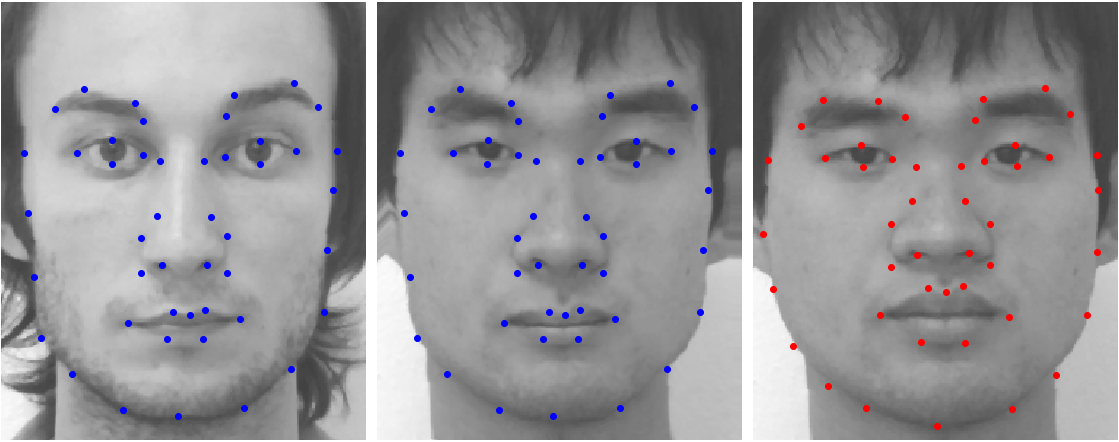}\\ 
  \caption{an example of usage of the software for image deformation. The image in the center is the image deformation of the right image with the landmarks corresponding to the left image.}
  \label{fig:facedeform}
\end{center}
\end{figure}

In fig.\ref{fig:facedeform} we illustrate the effect of the used image deformation algorithm on a picture of the FEI database.

\paragraph*{How many landmarks are too many?} The results presented in the main text lead to a picture of the origin of the efficiency of the decoupling code $\Rdue$. In substance, the above analysis suggests that, for the decoupling to be worth, {\it the number of landmarks should be low enough, relatively to the number of pixels}. In this situation, few landmarks, of the order of some tenths, require few information to be encoded and, at the same time, they imply a large texture information gap in since the resulting uniformed images are more compressible. Increasing the number of landmarks would increase their description length ${\cal G}_2$ and, at the same time, it would entail a lower and lower increase in the texture information gap ${\cal G}_1$, eventually leading to a negative overall gap $\cal G$. Roughly speaking, if the landmarks were too much given the image resolution, they would encode information regarding the details of the shape that is already present in the original images, and that does not further help compressing them when uniformed consequently. 

In the following two paragraphs we discuss the the role plaid by $d_{\rm s}$ and $d_{\rm t}$. On the one hand, the texture coordinates description length ${\cal G}_1$ increases with the number of pixels $d_{\rm t}$. The description length of the shape coordinates ${\cal G}_2=L_{{\cal L}_{\rm tr},p_{\rm s}}({\cal L}_{\rm tr})$ increases as well with the number of landmark coordinates $d_{\rm s}$.\footnote{Both dependences are slightly over-linear for lower values of $d_{\rm s}$ and $d_{\rm t}$. In the case of texture coordinates, this is the over-linear behaviour that we observed in figure \ref{fig:trainingevidences}. Analogously, adding landmark coordinates $d_{\rm s}'>d_{\rm s}$ will lead to a shape description length slightly larger than $(d'_{\rm s}/d_{\rm s}){\cal G}_2$, since a finer description of the facial shape contour will reveal information beyond what can be deduced deterministically (e.g., by interpolation) from the coarser description of each image's shape in terms of a lower number of landmarks.} In both cases, however, the dependence quickly becomes linear in $d$: respectively, when $d_{\rm t}$ approaches the finer resolution (in our dataset, $h=300$), and when the number of landmarks is so large that the average distance between landmarks becomes of the order of the image grid space unit (or much sooner).  

On the other hand, the texture information gap ${\cal G}_1$ itself increases {\it with the number of landmarks}.\footnote{\label{footnoteresolution} We do not present such an analysis since a systematic exploration is limited in a dataset with moderate number of landmarks. For $\nl$ of the order of a dozen, the resulting description length critically depends on {\it which landmarks} are considered. Furthermore, the image deformation algorithms (needing a minimally dense mesh of landmarks covering the non-uniform parts of the images) necessarily induce artefacts in the uniformed images for low $\nl$ and, consequently, in the description lengths. Rather than a numerical estimation in our dataset of limited $d_{\rm s}$, we find clearer the order-of-magnitude argument presented in this paragraph.}
However, when the average inter-landmark distance becomes of the order of the grid space unit, $d_{\rm s}$ has no longer impact on ${\cal G}_1$. Increasing the number of landmarks will not influence the texture information gap. 

The two above straightforward arguments motivate an order of magnitude estimation of the maximum shape dimension $\bar d_{\rm s}$ (twice the maximum number of landmarks) beyond which the uniformation is no longer worth, for the largest of the considered resolutions, $h_{\rm max}=300$. Such an estimation is based on neglecting the over-linear dependence of ${\cal G}_2$ on $d_{\rm s}$ and the dependence of ${\cal G}_1$ on $d_{\rm s}$. 

The decoupling is efficient in the extent to which the information gap is large, and larger than zero. Under the above hypotheses, $\bar d_{\rm s}$ is consequently the dimension by which one must multiply the shape description length per coordinate in order to reach ${\cal G}_1$ (see figure \ref{fig:trainingevidences} and {\it Likelihood and evidence of shape coordinates}): it is $\bar d_{\rm s} \simeq {\cal G}_1 d_{\rm s}/L_{{\cal L}_{\rm tr},p_{\rm s}}({\cal L}_{\rm tr})$, or $\bar d_{\rm s} \simeq 1700\pm 100$, roughly less than one thousand landmarks.

We notice, however that, under these crude hypotheses, the information gap monotonically decreases with $d_{\rm s}$. The more interesting estimation not of the maximum but of the {\it optimal} number of landmarks, that maximises the information gap for a fixed image resolution, would require to take into account the non-linear dependences on $d_{\rm s}$, neglected in the above argument (see footnote \ref{footnoteresolution}). We expect, in any case, that such an optimal value would be much lower than the above estimation for its maximum value, probably of the order of few hundred of landmarks, for $h\simeq 300$. 

\paragraph*{Likelihood and evidence of the normal distribution.} We report the well-known expression for the Bayesian evidence, and related formulae, of the normal distribution associated to the $p$-PCA representation with $p$ principal components. Given a dataset $\cal D$ composed by $N$ $d$-dimensional vectors, $p$-PCA induces a likelihood probability distribution which is the normal distribution (supposing null averages):

\begin{align}
\ln P(\D|\T,p)/(Nd) = -\frac{1}{2} \left[ \ln (2\pi) + \ln\det(C) +\text{tr}\left( C^{-1} \cdot \Sigma \right) \right]
\end{align}
where $\Sigma$ is the unbiased estimator of the correlation matrix of the data $\D$, and where the parameters $\T=C$ are the theoretical correlation matrix, which in $p$-PCA is subject to exhibit its $d-p$ lowest eigenvalues equal to a common noise-level value $v$. The maximum likelihood estimation $\T^*$ for $C$ and $v$ are: $C^*=U\Lambda U^\dag$ where $U$ is an orthogonal matrix whose top $p$ eigenvectors are those of $\Sigma$, and where the diagonal matrix $\Lambda$ contains the top $p$ eigenvalues of $\Sigma$, $\Lambda_{ii}=\lambda_i$ for $i\le p$, and the remaining $d-p$ diagonal elements equal to $\Lambda_{ii}=v_p$ for $i>p$, with $v_p=(d-p)^{-1}\sum_{j>p}\lambda_j$.

For completeness, we report the expressions for the description length, the empirical entropy and the Occam factor, making explicit the dependence on the number of principal components $p$: 
\begin{align}
L_p(\D) &= -\ln P_p(\D) - d\ln \epsilon = \\
&={\sf S}_p(\D|\T^*)+{\sf O}(\T^*) \\
{\sf S}_p(\D|\T) &=-\ln P_p(\D|\T)-d\ln (\epsilon) \\
\ln P_p(\D) &= \ln P_p(\D|\T^*) - {\sf O}_p(\T^*)
\end{align}
where, as mentioned in the main text, in these equations $\T^*$ refers to the maximum likelihood estimator. The equation for the Bayesian evidence (under certain assumptions on the prior variance) takes the form, up to a constant factor, and for sufficiently large $N$ \cite{minka2000}: 

\begin{align}
\ln P(\D)/(Nd) &\simeq \ln P(\D|\T)/(Nd) - \ln {\sf O}(\T^*)/(Nd) \\ 
-\ln {\sf O}(\T^*)/(Nd) &:= \frac{1}{2Nd}\left( (m+p+1) \ln(2 \pi) -p \ln N -\ln|A| + \ln |p_U| \right) \\
m &:= dp - p(p+1)/2 \\
\ln |p_U| &:= -p\ln 2 + \sum_{j=1}p \ln \Gamma\left(\frac{d-j+1}{2}\right) - \frac{d-j+1}{2}\ln\pi
\end{align}
and where:

\begin{align}
\ln |A| &= \sum_{i=1}^p \Bigg\{ \sum_{j=i+1}^d \left[ \ln(\hat\lambda_j^{-1}-\hat\lambda_i^{-1}) + \ln(\lambda_i-\lambda_j) \right] \Bigg\} + m \ln N 
\end{align}
where the $\lambda$'s are the eigenvalues of $\Sigma$ in decreasing order, $\hat\lambda_j=\lambda_j$ for $j\le p$ but $=v_p$ for $j>p$.

In the case $d>N$, this last term takes the form:

\begin{align}
\ln |A| &= \sum_{i=1}^p \Bigg\{ \sum_{j=i+1}^p \left[ \ln(\lambda_j^{-1}-\lambda_i^{-1}) + \ln(\lambda_i-\lambda_j) \right] + \\
&+ (d-p)\ln(v_p^{-1}-\lambda_i^{-1}) + \sum_{j=p+1}^N \ln(\lambda_i-\lambda_j) + \\
&+(d-N)  \ln\lambda_i \Bigg\} +m \ln N 
\end{align}
while for $d\le N$, it is:

\begin{align}
\ln |A| &= \sum_{i=1}^p \Bigg\{ \sum_{j=i+1}^p \left[ \ln(\lambda_j^{-1}-\lambda_i^{-1}) + \ln(\lambda_i-\lambda_j) \right] +  \\ 
&+ (d-p)\ln(v_p^{-1}-\lambda_i^{-1}) + \sum_{j=p+1}^d \ln(\lambda_i-\lambda_j)  \Bigg\}+m \ln N
.
\end{align}

\begin{figure}
\begin{center} 
  \includegraphics[width=12cm]{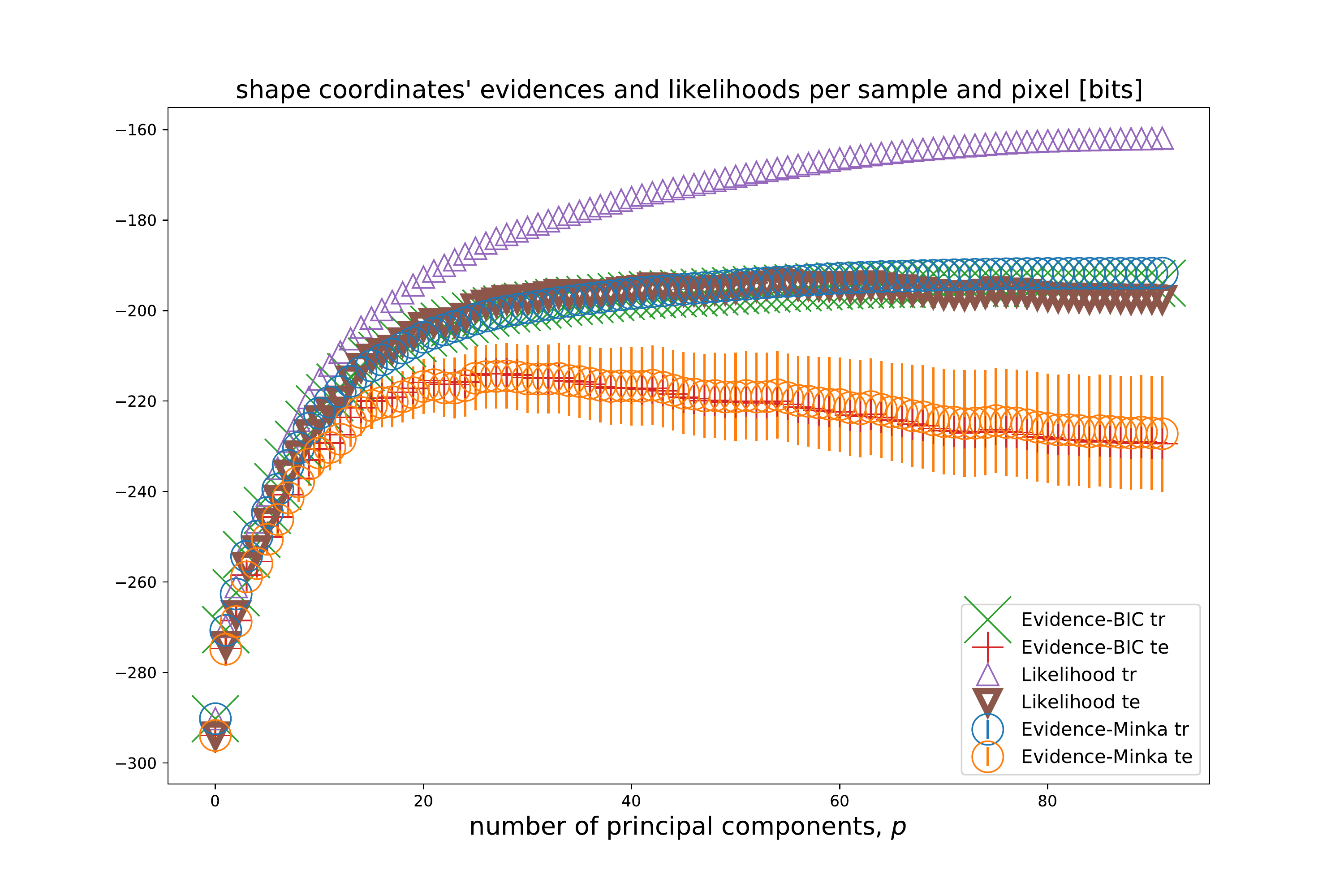}
  \includegraphics[width=12cm]{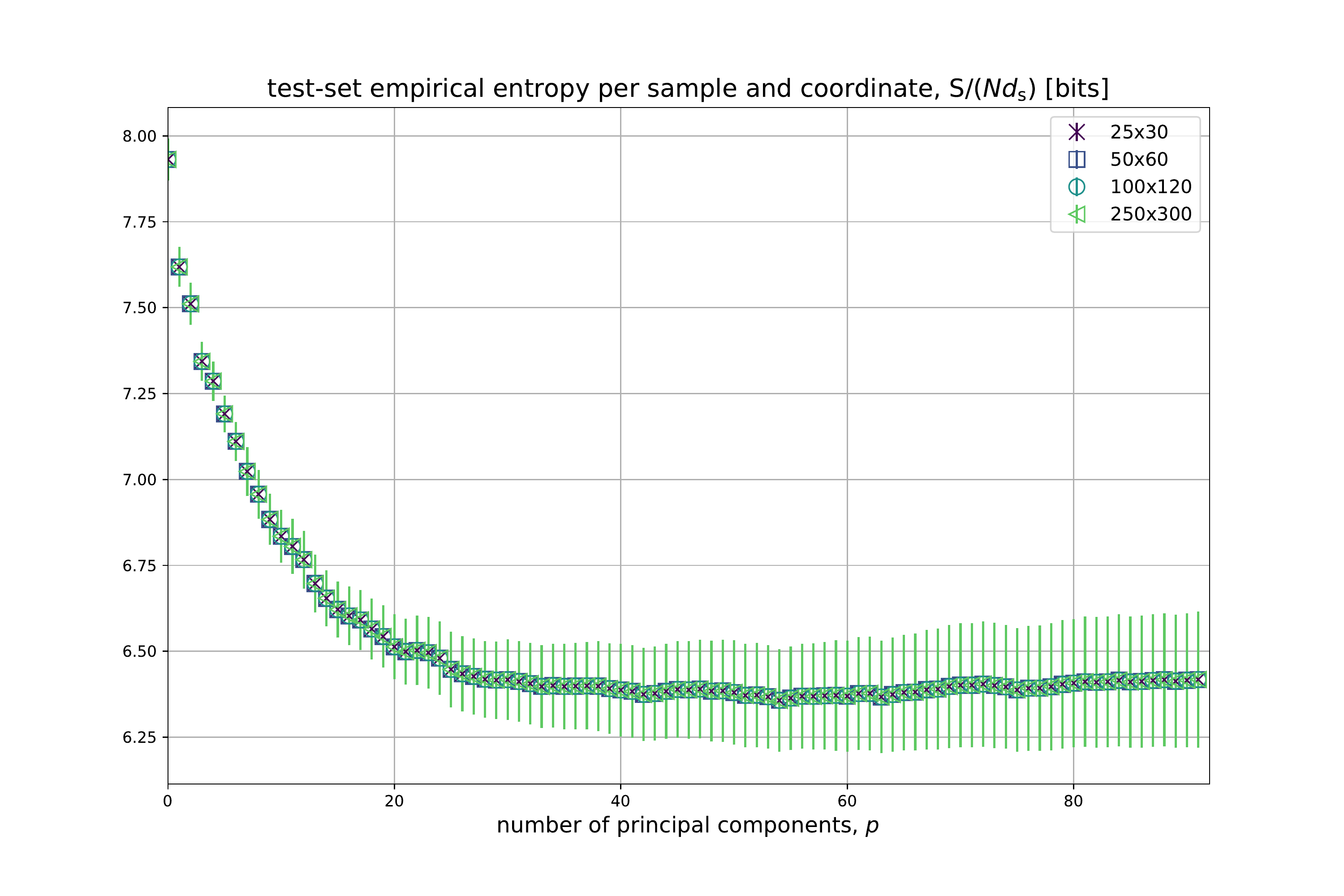}
  \caption{Left: Shape coordinates' likelihoods and evidences (in BIC approximation) of the test and training sets. Right: Empirical entropy of shape coordinates for the full $N_{\rm tr}=400$ training set, for different resolutions.}
  \label{fig:evidences_shape}
\end{center}
\end{figure}

\paragraph*{Likelihood and evidence of shape coordinates.} For shape coordinates, and for the datasets considered here, it is $d_{\rm s}<N$. In figure \ref{fig:evidences_shape} (upper panel) we show the behaviour of the training- and test-set (logarithms of the) likelihood, along with the training- and test-set (logarithms of the) Bayesian evidence of shape coordinates (respectively, $\ln P({\cal L}_{\text{tr}}|C^{{\rm s}})$, $\ln P({\cal L}_{\text{te}}|C^{{\rm s}})$, $\ln P({\cal L}_{\text{tr}})$, $\ln P({\cal L}_{\text{te}})$). We observe that the training-set evidence behaviour is qualitatively similar to that of the the test-set likelihood (contrary to the case of texture coordinates, see below). 

When commenting the results of figure \ref{fig:gap_info_vs_resolution}, we mentioned the fact that the empirical entropy of shape coordinates does not depend on the resolution. Indeed, changing the resolution in the dataset of shape coordinates amounts to multiply the Landmarks' Cartesian coordinates by a factor ($w/w'$ for horizontal, $h/h'$ for vertical coordinates). However, the relevant quantity in these experiments is not the absolute value of the coordinates in the $w\times h$ grid units, but their normalised value in units of the image heigh $h$. If normalised coordinates are considered, the precision should be consequently normalised to be inversely proportional to $h$. In figure \ref{fig:evidences_shape} (lower panel) we plot the training empirical entropy $S_p({\cal L}_\text{tr}|C^{{\rm s}})=-\ln P({\cal L}_\text{tr}|C^{{\rm s}})-d_{\rm s}\ln \epsilon_{\rm s}$ for different resolutions, using the resolution-dependent precision $\epsilon_{\rm s} = 0.1 (h_\text{max}/h)$. The overlap of different curves is a consequence of the fact that no information has been lost when scaling both the coordinates and the precision.

\begin{figure}
\begin{center} 
  \includegraphics[width=12cm]{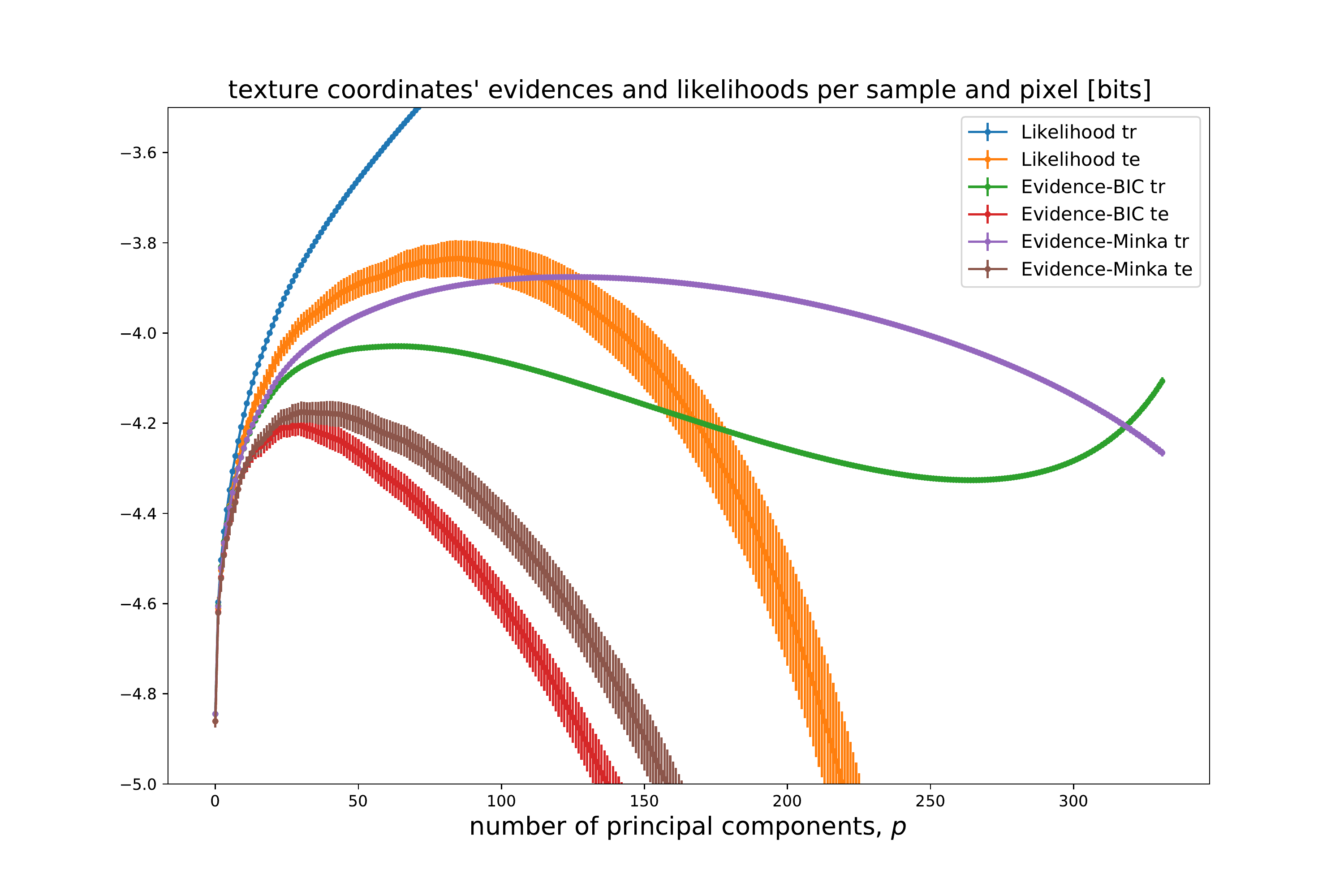}
  \caption{Uniformed texture coordinates' likelihoods and evidences of the test and training sets.}
  \label{fig:textures_shape}
\end{center}
\end{figure}

\paragraph*{Likelihood and evidence of texture coordinates.} In figure \ref{fig:textures_shape}  we show the behaviour of the training- and test-set (logarithms of the) likelihoods along with the training- and test-set (logarithms of the) Bayesian evidence of texture coordinates. In this case, in which, differently from shape coordinates, it is $d_{\rm t}\gg N$, the BIC approximation for the evidence is, as expected, no longer good. Moreover, the evidence behaves differently from the test-set likelihood. In order to perform model selection in this case, or to estimate the Occam contribution to the description length, it is necessary to use the aforementioned expression of the Bayesian evidence due to Minka.

The reader may wander why the texture coordinates' seem to present strong overfitting in figure \ref{fig:textures_shape}, where the test-likelihood presents a fast decreasing for $p\gtrsim 100$, while the opposite seem to hold in the facial recognition task, figure \ref{fig:recognition}, where the shape coordinates present a minimum of the classification error rate for $p\gtrsim 30$, while texture coordinates do not seem to overfit in this case. On the one hand, in the case of figure \ref{fig:textures_shape}, this is what one expects in a situation in which $d\gg N$. In this situation, the training-set variance of the $p$-th modes $\lambda_p$ are underestimated (see, for example \cite{bun2017}) and, consequently,  the negative test-set {\it energy} (or minus the exponent of the normal distribution probability density, $\traza(C_{\rm te}\cdot C_{\rm tr}^{-1})=\sum_k^p{\<x'_k}^2\>_{\rm te}/\lambda_k$) increases fast with $p$. This is the difference between the test and training log-likelihood. Such a difference is also present for shape coordinates in figure \ref{fig:evidences_shape}, in this case the training-test difference in likelihood is lower (and roughly equal to $-m\ln(N)/2N$, since Minka and BIC evidences coincide), meaning that also in this case $x'_k$ varies more than $\lambda_k^{1/2}$ in the test-set. 

On the other hand, for the classification error, the (squared) Mahalanobis similarity between the $s$-th and $s'$-th subjects is:

\begin{align}
\sum_{j=1}^p \frac{1}{\lambda_j} \left(y'_j(s\text{\small , smile})-y'_j(s'\text{\small , neutral})\right)^2
\end{align}

The difference, in this case, is that shape coordinates vary drastically between smile and neutral facial images, and are not that characteristic of the subject's identity as texture coordinates are. The high-$j$ terms in the above equation present strong random fluctuations even if $s'=s$, since high-$j$ fluctuations $y'_j$ are not representative of the subject, and are over-weighted by the relatively smaller $\lambda_j$. This is the origin of the increasing of the error rate with $p$. The same over-weighting of the fluctuations happens for texture coordinates, although, in this case, $D_{s,s',j}=y'_j(s\text{\small , smile})-y'_j(s'\text{\small , neutral})$ is still more similar when $s'=s$ than for $s'\ne s$ even for hight $j$, {\it and/or} $D_{s,s',j}$ is so low for low values of $j$, that it compensates the overall increase in $D_{s,s',j}$ for larger values of $j$. The fact that the error rate does not seem to increase at all with $p$ in figure \ref{fig:recognition} suggests that the first of the two hypothesis in the above sentence holds. 

\begin{figure}
\begin{center} 
  \includegraphics[width=5cm]{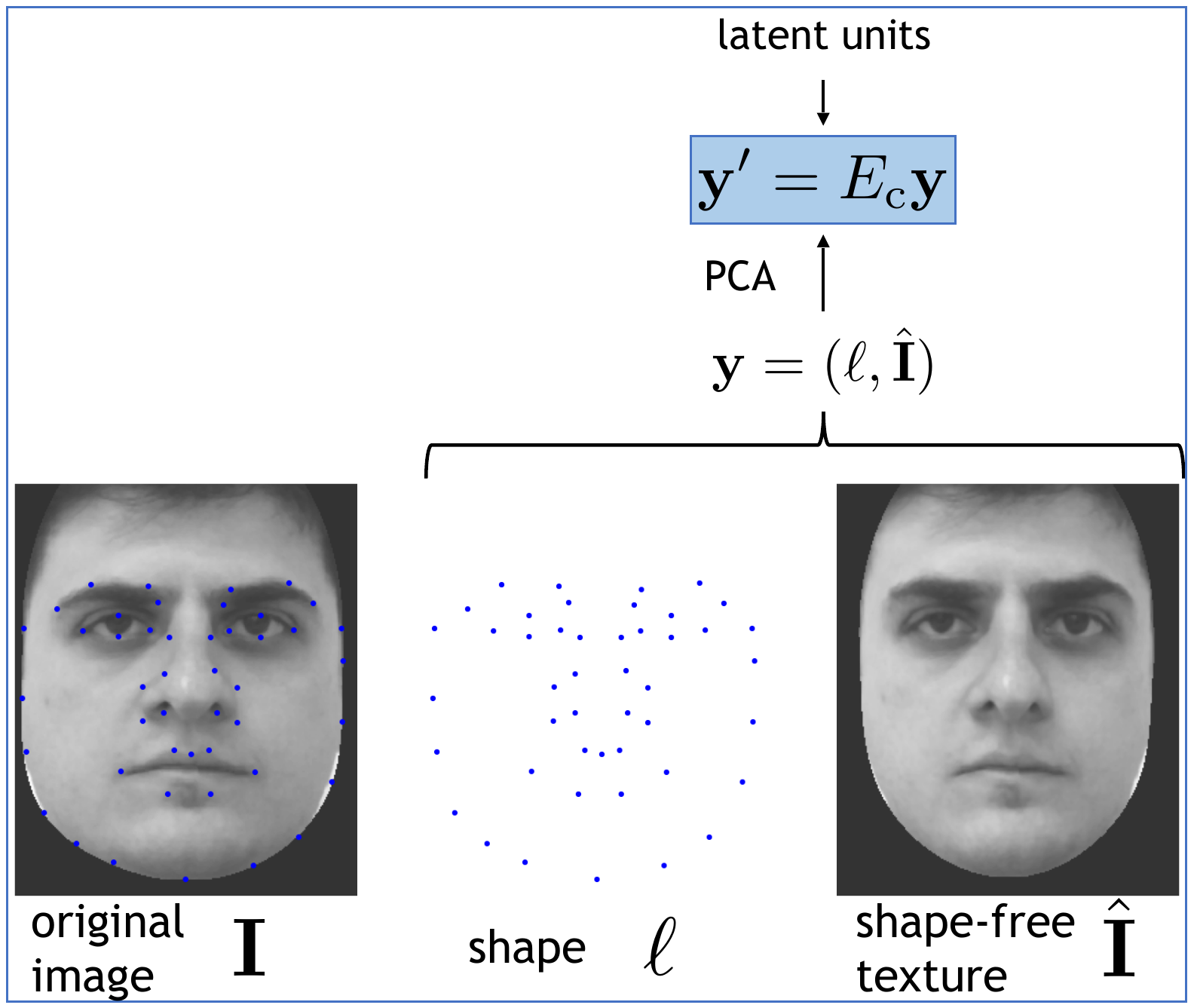}
  \caption{Schematic representation of the concatenated code.}
  \label{fig:concatenated_code}
\end{center}
\end{figure}

\paragraph*{The concatenated code representation, $\Rtre$} (see figure \ref{fig:concatenated_code}), simply consists in concatenating the uniformed texture and shape coordinates in a single vector ${\bf y}=({\bm \ell},\hat\I)$, and to keep the first $p_\text{c}$ principal components of the set of concatenated vectors ${\bf y}'=E^{(\text{c})}{\bf y}$, hence treating shape and texture coordinates on the same ground. For an image dataset such that shape and texture coordinates were completely uncorrelated (say, $\<\ell_m I_i\>=0$ $\forall m,i$), the concatenated code would exactly coincide with $\Rdue$, in the sense that each principal axis would be a (normalised) concatenation of principal axes of ${\bm \ell}$ and $\hat\I$ coordinates. The performance of the $\Rtre$ code in the face processing tasks presented in section Results turns to be almost identical using texture coordinates only. The reason is that shape coordinates carry a lower amount of aggregated information and, in any case, the correlations between shape and texture coordinates are significantly smaller than those in the diagonal blocks of $C^{(\text{c})}$. The advantage of using $\Rtre$ is that one may fix a single number of principal components. The daydream generation of novel facial images with the $\Rdue$ code (fixing $p_{\rm s}=d_{\rm s}$ at its maximum value) leads to almost identical results of those of $\Rtre$ in figure \ref{fig:sampling}.

\paragraph*{Details of the classification algorithms.} The classification tasks are performed via a nearest-neighbour classifier: every vector $\bf{x}$ is assigned to the class that minimizes the distance from $\bf{x}$. If a class contains more than one element, as it is the case of the gender classification task (in which the male and the female classes contain $200$ vectors each, corresponding to half of the raw FEI database), the distance is computed between $\bf{x}$ and the average of the elements belonging to the class.
For the gender identification task we follow a leave-one-out approach: for each vector $\bf{x}$, the training-set (from which we compute the correlation matrix, defining in its turn the Mahalanobis distance ${\sf d}_p(\cdot,\cdot)$) is composed by all the dataset vectors except for $\bf{x}$ itself. The so defined training-set is as well the set from which the average vector of each class is constructed. 

In the facial recognition task, we use a more economic strategy: we construct $K=5$ training/test-set divisions (with $N_\text{te}=400/K=80$, $N_\text{tr}=320$) by $K$-folding, in such a way that the test-set contains at most one vector per individual, and that it contains $N_\text{te}/2$ vectors corresponding to smiling portraits, and $N_\text{te}/2$ corresponding to neutral portraits. For each of the vectors of facial coordinates in the test-set, we search for its nearest neighbor among the $N_\text{tr}=320$ vectors of facial coordinates in the training set. The training set is, again, the set from which we compute $C_p$ and consequently define ${\sf d}_p(\cdot,\cdot)$. Afterwards, the average value and the standard deviation of the mean of the success rate is computed by cross-validation over the $K=5$ iterations. 

\begin{figure}[]
\begin{subfigure}{0.48\textwidth}
\includegraphics[width=\linewidth]{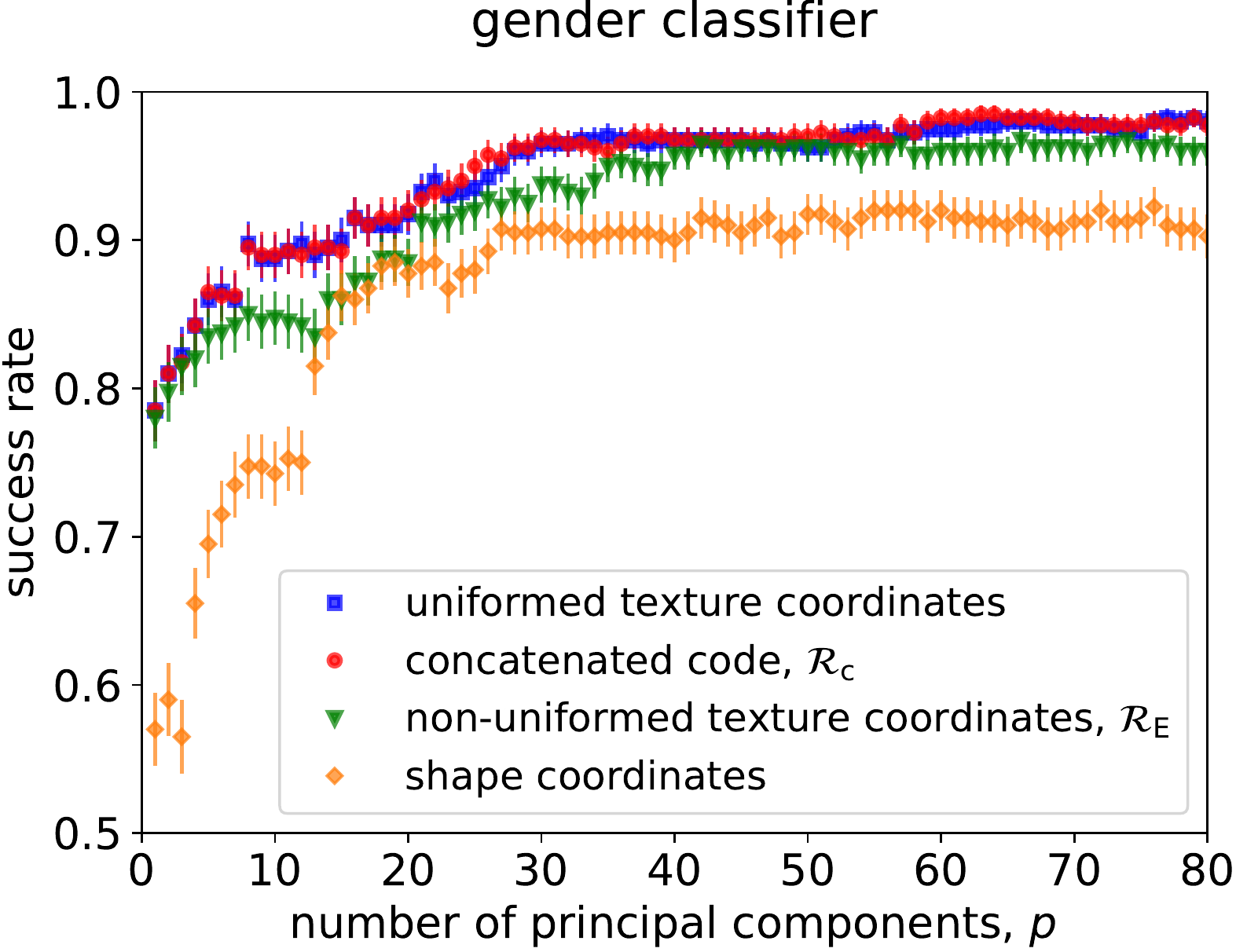}
\caption{}
\label{subfig:gender_far}
\end{subfigure}
\quad
\begin{subfigure}{0.48\textwidth}
\includegraphics[width=\linewidth]{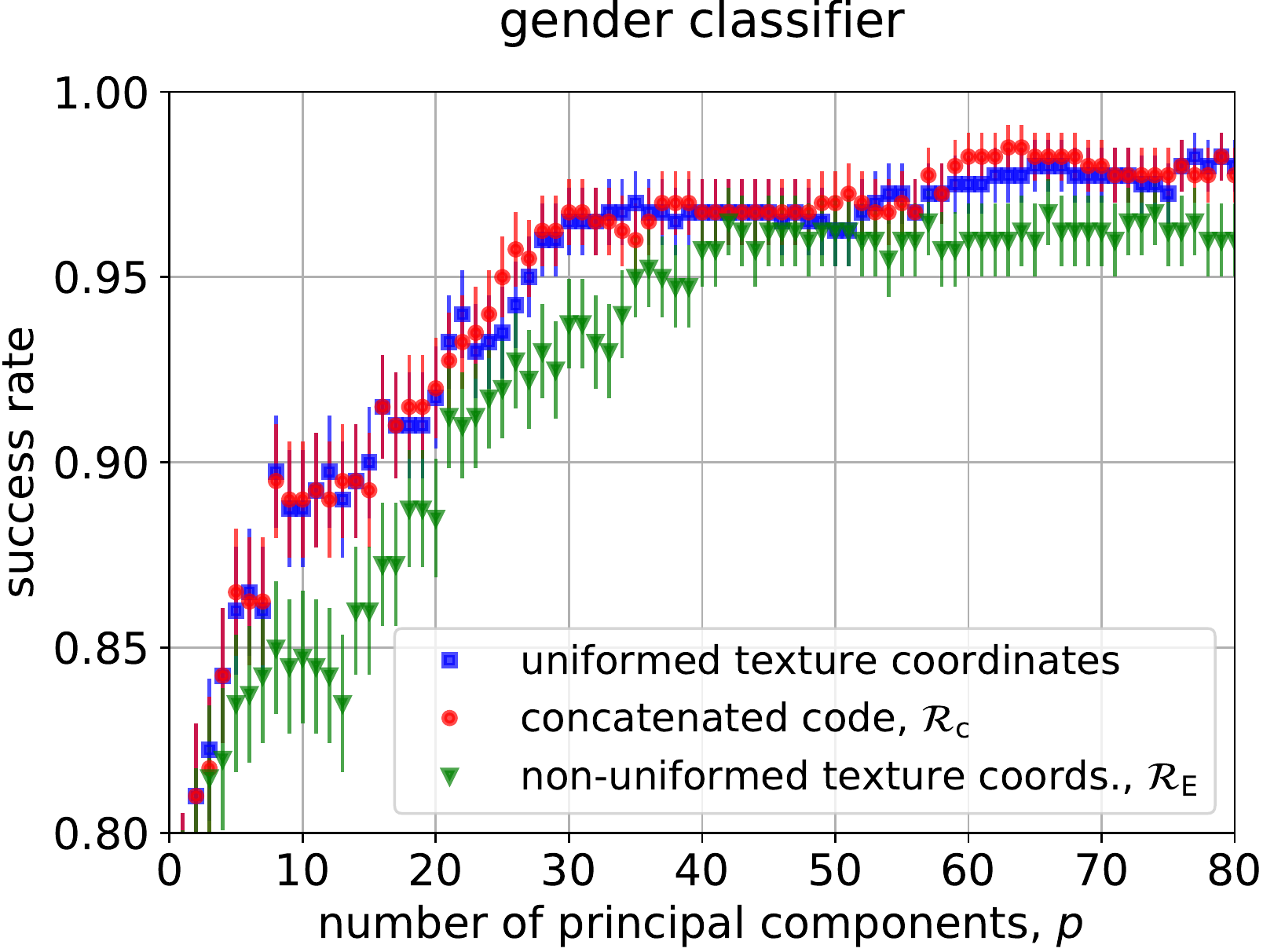}
\caption{}
\label{subfig:gender_close}
\end{subfigure}
\caption{the success rates in the tasks of gender classification and face recognition are here displayed -on the left- as functions of the number $P$ of principal components; in the right column the same data (except for the geometric coordinates) are shown in a close-up.} 
\label{fig:genderclassification}
\end{figure}

\paragraph*{Results of the gender classification task.} In figure \ref{fig:genderclassification} we present the results of the gender classification task. We observe that the shape coordinates alone are sufficient to achieve roughly $90\%$ of successful attempts with less than $30$ PC's. Consistently with the rest of the article results, the classification performed in terms of (principal components of) uniformed facial images achieves higher success rates respect to that using (principal components of) the original original facial image (i.e., the $\Runo$ representation). Furthermore, the success rate plateau is reached for a lower number of PC's ($p\simeq30$ versus $p\simeq40$ of $\Runo$).

\paragraph*{Different regularisation schemes.} For each generic set of facial coordinates (say, ${\cal D}$), we have so far estimated its description length according to the $p$-PCA model, whose number of principal components $p$ are those that maximise the description length $p^*=\arg \min_p L_{{\cal D},p}({\cal D})$. The inferred probability distribution is a normal distribution whose correlation matrix $C_{p^*}$ is consequently different from the empirical correlation matrix, say $C_{p=\min\{N,d\}}$, since not all empirical eigenvalues and eigenvectors are statistically significant given the dataset finiteness.\footnote{Strictly speaking, in the $N<d$ case, the inferred correlation matrix $C_p$ is different from the empirical matrix even if $p=N$, since it has to be regularised so that its rank is $=d$ (and not $=N$).} The normal distribution whose correlation is the empirical matrix would correspond, instead, to maximum likelihood inference.

Actually, there are different ways, besides $p$-PCA, in which the correlation matrix may be inferred beyond the maximum likelihood criterion. An alternative is {\it linear (identity) shrinkage} (see, for example, \cite{bun2017}). Linear shrinkage leads to a correlation matrix which is a convex combination between the unbiased (maximum likelihood) empirical estimator $C$ and a completely biased (and null-variance) matrix, as the identity matrix in $d$ dimensions $\1_d$. In other words, the ``regularised'' shrunk matrix is $C_\alpha=(1-\alpha) C+\alpha\1_d$ where $\alpha$ is a real number in $[0,1]$, that may be chosen by maximum (cross-validated) out-of-sample likelihood. In the $p$-PCA scheme, $p=0$ and $p=\min\{N,d\}$ are the arguments of the minimum and maximum training likelihood respectively, and $p^*$ is comprised between them. Within the shrinkage scheme, these extreme cases correspond to $\alpha=1$ and $0$, respectively.

\begin{figure}
\begin{center}
    \makebox[\textwidth][c]{\includegraphics[width=\textwidth]{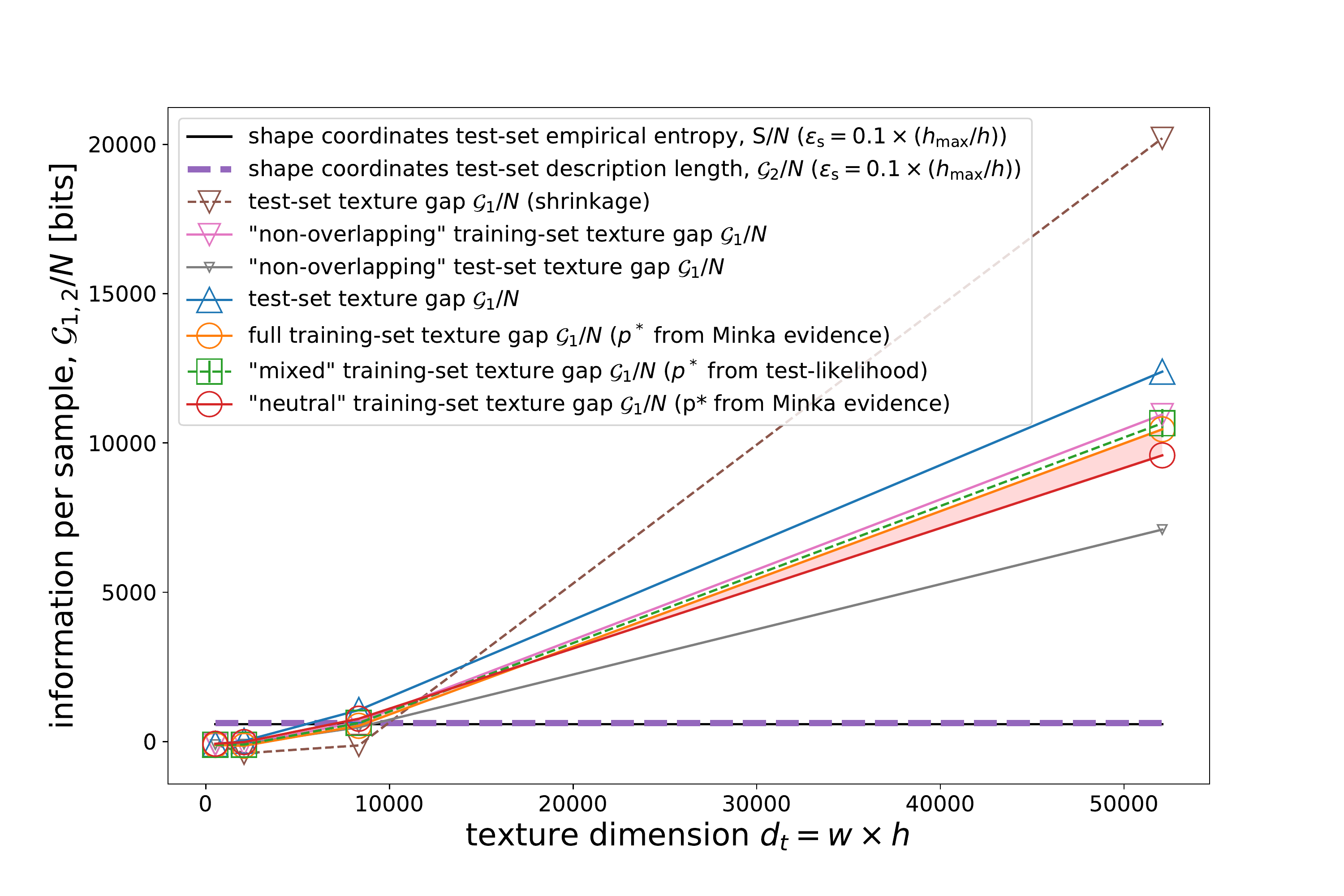}}
	\caption{The same as figures \ref{fig:gap_info_vs_resolution},\ref{fig:gap_info_vs_resolution_test} but with the addition of the test-set texture gap per sample ${\cal G}_1/N$ computed with the shrinkage regularisation method.}
    \label{fig:gap_info_vs_resolution_shrinkage}
\end{center}   
\end{figure}

In order to check the robustness of our results with respect to the regularisation scheme, we have computed the information gaps (actually, the gaps in empirical entropy)\footnote{We make notice that, in the case of $p$-PCA, and for texture coordinates, the texture gap ${\cal G}_1$ is essentially given by the gap between empirical entropies. The difference between the Occam factors of non-uniformed and uniformed images is negligible in front of it.} resulting from the normal probability distributions associated not with $p$-PCA but with linear shrinkage. We have observed that the results are qualitatively consistent with those presented here. While the lowest description length of the set of landmarks ${\cal G}_2$ is consistent with the one shown in figure \ref{fig:gap_info_vs_resolution}, the description length gap ${\cal G}_1$ is significantly larger for the largest resolution, as can be seen in figure \ref{fig:gap_info_vs_resolution_shrinkage}. Consequently, the information gap is even larger when regularising the correlation matrices with the shrinkage method.

\begin{figure}
\begin{center}
    \includegraphics[width=\textwidth]{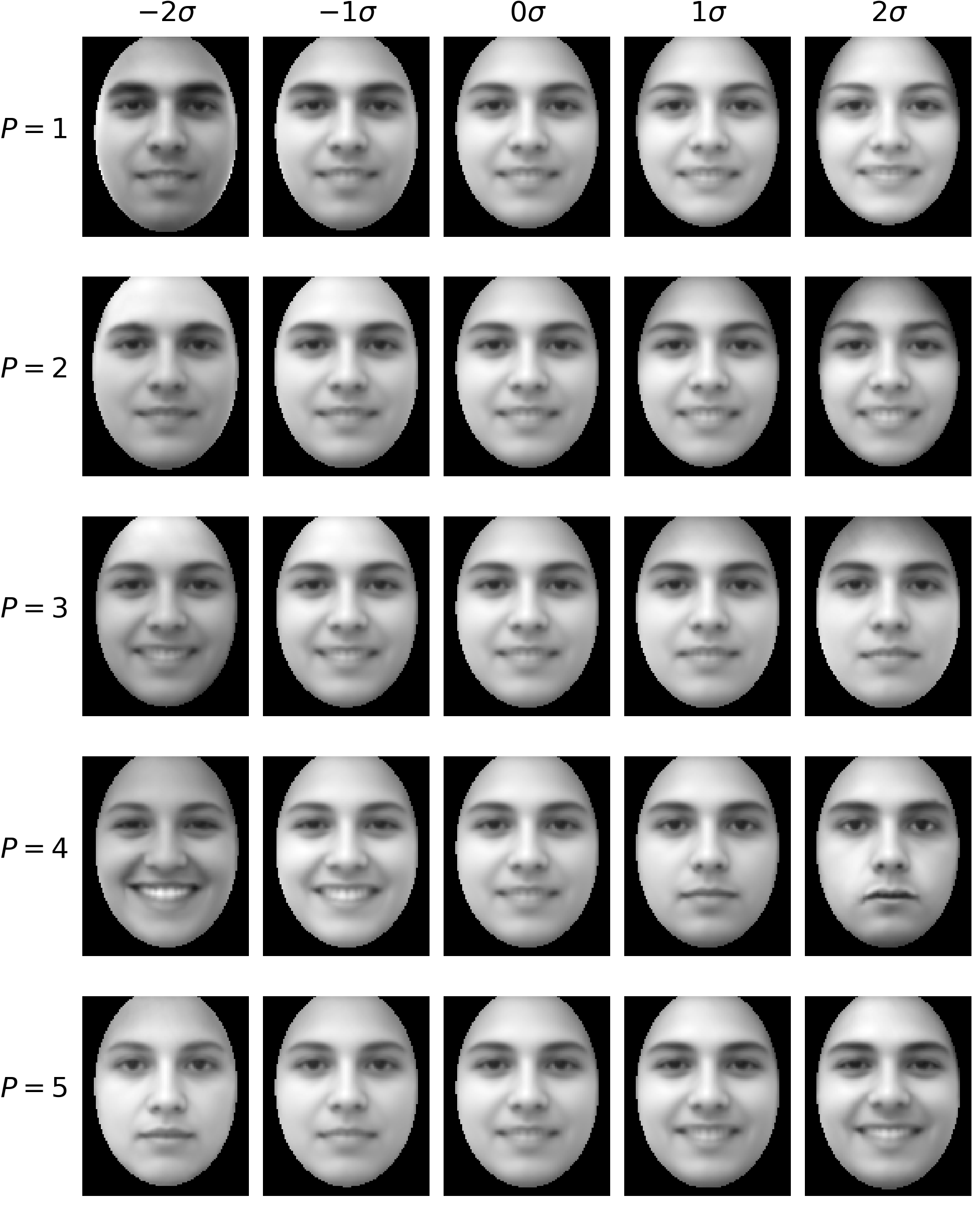}
    \caption{First five principal axes of the concatenated code $\Rtre$ (five largest-eigenvalue eigenvectors of $C^{(\text{c})}$). The $j$-th column represents the $j=th$ eigenvector. In particular,  the $i$-th row of the table represents the points that have all the coordinates,in the base of the principal axes, equal to zero except the $i$-th one, that ranges from $-2\sigma$ (left) to $-2\sigma$ (right); $\sigma$ is taken equal to the square root of the largest eigenvalue $\lambda_1$ of the correlation matrix. In other words, the image, say $\I$ in the $i$-th row, $j$-th column is obtained by de-uniformation $(\<\bm \ell\>,\hat\I) \to ({\bm \ell},\I)$, where ${\bm \ell}$ and $\hat \I$ are obtained as: $({\bm\ell,\hat\I})={\bf y}=E^\dag\cdot {\bf y}'$, and where ${\bf y}'$ is the vector that exhibits null principal components except by the $i$-th, $y'_i=(j-3)\lambda_1^{1/2}$, and $E$ is the matrix of row eigenvectors of $C^{(\text{c})}$.} 
\label{fig:principal_axes_nc}
\end{center}   
\end{figure}

\paragraph*{Visualisation of the eigenvectors of the concatenated code $\Rtre$.} Figure \ref{fig:principal_axes_nc} presents a graphical visualisation of the first five principal axes of the whole database according to the concatenated code $\Rtre$ (the first five eigenvectors of $C^{(\text{c})}$).

\section{Acknowledgments}

We thank Matteo Marsili for his comments on the manuscript. This research received funding from the European Union’s Horizon 2020 Framework Programme for Research and Innovation under the Specific Grant Agreement No. 945539 (Human Brain Project SGA3) to GP and the European Research Council under the Grant Agreement No. 820213 (ThinkAhead) to GP. M. I.-B. is supported by the grant EU FESR-FSE PON Ricerca e Innovazione 2014-2020 BraVi.

\newpage
\section*{References}

\bibliography{face_bibliography}
\end{document}